\documentclass[12pt]{article}
\usepackage[margin=1.25in]{geometry}
\usepackage{graphicx,  amssymb, array, amsmath, float}
\usepackage{setspace}
\usepackage{color}
\usepackage[authoryear]{natbib}
\usepackage{bm}
\usepackage{graphicx}
\usepackage{float}
\usepackage{multicol}
\usepackage{xr}
\usepackage{mathtools}
\usepackage{comment}
\usepackage{nameref}
\usepackage{hyperref}
\usepackage{subfig}
\usepackage[ruled]{algorithm2e}

\newcommand{\iid}{\stackrel{\mathrm{iid}}{\sim}}

\begin{document}

\title{\large \bf Risk-Efficient Bayesian Data Synthesis for Privacy Protection}

\author{Jingchen Hu\footnote{Vassar College, Box 27, 124 Raymond Ave, Poughkeepsie, NY 12604, United States, jihu@vassar.edu} $\,$ and Terrance D. Savitsky\footnote{U.S. Bureau of Labor Statistics, Office of Survey Methods Research, Suite 5930, 2 Massachusetts Ave NE Washington, DC 20212, United States, Savitsky.Terrance@bls.gov} and Matthew R. Williams\footnote{National Center for Science and Engineering Statistics, National Science Foundation, 2415 Eisenhower Avenue, Alexandria, VA 22314, United States, mrwillia@nsf.gov}}
\maketitle

\begin{abstract}

Statistical agencies utilize models to synthesize respondent-level data for release to the public for privacy protection. In this work, we efficiently induce privacy protection into any Bayesian synthesis model by employing a pseudo likelihood that exponentiates each likelihood contribution by an observation record-indexed weight $\in [0, 1]$, defined to be inversely proportional to the identification risk for that record.  We start with the marginal probability of identification risk for a record, which is composed as the probability that the identity of the record may be disclosed. Our application to the Consumer Expenditure Surveys (CE) of the U.S. Bureau of Labor Statistics demonstrates that the marginally risk-weighted synthesizer provides an overall improved privacy protection. However, the identification risks actually increase for some moderate-risk records after risk-weighted pseudo posterior estimation synthesis due to increased isolation after weighting, a phenomenon we label ``whack-a-mole." We proceed to construct a weight for each record from a collection of \emph{pairwise} identification risk probabilities with other records, where each pairwise probability measures the joint probability of re-identification of the pair of records, which mitigates the whack-a-mole issue and produces a more efficient set of synthetic data with lower risk and higher utility for the CE data.

{\bf{keywords}}: Bayesian hierarchical models, Data privacy protection, Identification risks, Pairwise, Pseudo posterior, Synthetic data

\end{abstract}

\section{Introduction}
\label{intro}
Statistical agencies collect respondent-level data, also known as microdata, from households and business establishments through survey and census instruments. Researchers and data analysts often seek access to the respondent-level, detailed data records from statistical agencies in order to facilitate their research inferential goals. For example, researchers and policymakers could conduct regression analyses using variables in microdata.  Indeed, statistical agencies often disseminate public use microdata files to facilitate such purposes. Findings from these analysis projects, in turn, help policymakers to make data-driven decisions. In short, there is a substantial benefit from disseminating microdata to the public by statistical agencies.

When disseminating public use microdata files, however, statistical agencies are under legal obligation to protect privacy and confidentiality of respondents (e.g. U.S. Title 13). Therefore, the collected microdata has to undergo statistical disclosure control (SDC) procedures before its public release. SDC seeks to reduce the risk that a potential intruder is able to uncover the identity of or infer sensitive information about any data record. SDC procedures for microdata include adding random noise, topcoding, and swapping. In addition to microdata, statistical agencies publish summary statistics and tables to the public, which also require SDC procedures.  We refer interested readers to \cite{SDC2012book} for a comprehensive overview of SDC.



For microdata dissemination with privacy protection, generating and releasing synthetic data is an SDC approach that has gained popularity. Synthetic data is released in place of the confidential data and encodes privacy protection by performing smoothing of the original data distribution \citep{Rubin1993synthetic, Little1993synthetic}. Statistical agencies develop Bayesian models, called ``synthesizers," and apply them to the confidential data. They simulate records from the posterior predictive distributions of the estimated models and release the synthetic microdata to the public. Well designed synthesizers tailored to specific data types and features are able to maintain high utility.  A high utility signifies that data users can obtain inference results from the synthetic data that closely resemble those from the confidential data. 
 
From the privacy protection perspective, synthesizers encode privacy protection by smoothing the confidential data distribution.  Smoothing often softens or removes ``local" features from the confidential data distribution, induced by relatively few records.   Such local features suggest that the values for these records are relatively ``isolated" as compared to a global feature, such as a mode surrounded by many data records.  The identities of isolated records are more easily discovered unless those local features are reduced or eliminated.
Recent examples of Bayesian synthesizers include: \citet{ QuickHolanWikle2018JRSSA} and \citet{DrechslerHu2018} proposed methods to preserve spatial features of confidential data; \citet{WeiReiter2016} proposed methods for business establishment surveys with fixed total constraints; \citet{ManriqueHu2018JRSSA} proposed algorithms to deal with structural zeros (e.g., impossible combinations); and \citet{HuReiterWang2018BA} proposed methods for individuals nested in households.

A number of statistical agencies in several countries have created public use synthetic microdata products, including the synthetic Longitudinal Business Database \citep{SynLBD2011} 
by the U.S. Census Bureau, the IAB Establishment Panel \citep{DrechslerDundlerBenderRasslerZwick2008} in Germany, and synthetic business microdata disseminated by the Canadian Research Data Centre Network, among others. Moreover, depending on the protection goals, statistical agencies can choose between fully synthetic data, where all variables are deemed sensitive and synthesized \citep{Rubin1993synthetic}, and partially synthetic data, where only a subset variables is deemed sensitive and synthesized \citep{Little1993synthetic}. We refer interested readers to \citet{Drechsler2011book} for details of fully and partially synthetic data, the necessity of generating and releasing multiple synthetic datasets, and combining rules for valid inferences.

Utility evaluation of synthetic data is usually two-fold. Analysis-specific utility measures evaluate how closely inferential results obtained on the synthetic data resemble those on the confidential data. For example, \citet{KarrKohnenOganianReiterSanil2006} propose a confidence interval overlap measure to evaluate how much confidence intervals of regression coefficients from the confidential and synthetic data overlap. Global utility measures aim at evaluating the closeness between the confidential distribution and the synthetic distribution. 
Commonly used global utility measures include propensity scores and empirical CDFs \citep{Woo2009JPC, Snoke2018JRSSA}.

Disclosure risk evaluation of synthetic data focuses on its level of privacy protection. The two most common types of disclosure risks are: identification disclosure, which quantifies the probability of discovering the identity of a record, and attribute disclosure, which quantifies the probability of inferring sensitive information of a record \citep{Hu2019TDP}. Evaluation of these disclosure risks relies on assumptions of intruder's knowledge and behavior. In this work, we focus on synthetic data and measure the identification risk for each data record by directly computing a probability of identification disclosure ($\in [0,1]$), an extension of the expected match risk measure of \citet{ReiterMitra2009}.
Computing this probability, however, requires mild assumptions about the behavior of the intruder that, while intuitive, could be unverifiable, in practice.

Differential privacy, by contrast, constructs mechanisms that are guaranteed to produce a chosen, measured level of privacy protection that does not depend on any assumptions about the behavior of the intruder 
\citep{Dwork:2006:CNS:2180286.2180305}.  \citet{Dimitrakakis:2017:DPB:3122009.3122020} define a posterior mechanism for synthesizing data that generally requires truncation of the space of parameters to achieve a privacy guarantee. We seek to avoid this truncation of parameters by using our probability of identification disclosure as an alternative measure of risk or privacy protection.

In this work, we propose a general framework which leverages the pseudo posterior to provide a larger degree of privacy protection for high-risk records in the confidential data. We first evaluate the identification risk probability of \emph{each} record in the \emph{confidential} data, denoted as $IR_i^c$ for record $i \in (1, \cdots, n)$, where $c$ denotes confidential data. The $IR_i^c$ is a \emph{marginal} probability of identification risk for record $i$, and the closer the $IR_i^c$ value to 1, the higher the identification risk (i.e. probability of being identified) of record $i$. Second, we design a record-indexed weight $\alpha_i \in [0, 1]$, based on $IR_i^c$, which is inversely proportional to $IR_i^c$ and bounded between 0 and 1. We introduce the collection of vector weights, $\bm{\alpha} = (\alpha_i \in [0, 1])_{i=1, \cdots, n}$, to exponentiate the likelihood contributions in a pseudo likelihood framework that, when convolved with the prior distributions, produce a joint pseudo posterior distribution. See \citet{2015arXiv150707050S} for background on a pseudo posterior distribution constructed in the case of complex survey sampling. Owners of confidential data are then enabled to generate synthetic microdata from the pseudo posterior predictive distribution. This construction surgically or locally \emph{downweights} the likelihood contributions for those records that express high risks (high $IR_i^c$ produces low $\alpha_i$). By surgically downweighting likelihood contributions for riskier records, 
the risk-weighted pseudo posterior is designed to produce an ``efficient" reduction in disclosure risk, calculated as by-record marginal disclosure risk probability in the \emph{synthetic} data and denoted by $IR_i$, for data record, $i \in (1,\ldots,n)$.  
We refer to the selective downweighting of high risk records as ``surgical" or precise.  A relatively more efficient synthesizer induces a smaller loss of data utility in order to achieve a targeted disclosure risk profile.

We use the proposed risk-weighted synthesizer based on the marginal downweighting approach to synthesize a highly-skewed and sensitive family income variable in a sample collected for the Consumer Expenditure Surveys (CE) at the U.S. Bureau of Labor Statistics (BLS). Our approach is offered as a replacement for the current BLS practice of topcoding the family income variable to reduce the relative isolation of data values located in the distribution tail.  Topcoding refers to the practice of using a pre-chosen value and censoring any values above the pre-chosen topcoded value to that value \citep{AnLittle2007JRSSA}.  We demonstrate in the sequel that topcoding is blunt in that it eliminates the tails of the confidential data distribution, which in turn eliminates the ability to estimate non-central quantiles. We also demonstrate that there may be many records outside of the tails that express high risks. These high-risk records are left untouched by topcoding.  Our risk-weighted pseudo posterior synthesizer, however, reduces risk in all high-risk data records. In addition to the CE application, we demonstrate our downweighting scheme in a simulation study.

The use of the marginal probability to downweight the likelihood contributions may unintentionally increase the identification disclosure risks for some records with moderate risk in the confidential data by shrinking away nearby high risk records.  We refer to this issue 
as ``whack-a-mole." 
The use of marginal weights may overly shrink the tail or extreme observations and increase the relative isolation of a moderate-risk data record by further isolating it, thus increasing the $IR_{i}$ of that record in the generated synthetic data. To alleviate this issue and better control the utility-risk trade-off of the synthesizers, we further propose to formulate the weight for each record by constructing a collection of joint, \emph{pairwise} probabilities of identification risk for that record with the other data records. Our use of pairwise identification risk probabilities may be viewed as an adaptation of \citet{WilliamsSavitsky_pairwise2018} from the survey sampling case (where the weights are based on unit inclusion probabilities into a sample of a finite population).

The use of pairwise identification risk probabilities for formulating by-record weights ties together the downweighting of records. We show that it not only mitigates the whack-a-mole problem on the CE application, but also improves the utility preservation as compared to marginal downweighting. In addition, we construct and illustrate practical approaches to adjusting the weights to achieve a desired synthetic data utility-risk trade-off, included in the Supplementary Materials.


Section \ref{intro:CEdata} introduces details of the CE sample data in our application and their current topcoding practice in the CE public-use microdata (PUMD) undergone SDC. 

\subsection{The CE data and the topcoded family income}
\label{intro:CEdata}

The CE data sample comes from 2017 1st quarter, which contains n = 6208 consumer units (CU). A CU denotes either a household or independent entities within a household (such as roommates). We focus on 11 variables: the first 10 variables are categorical, considered insensitive, therefore, not to be synthesized and used as predictors\footnote{The age variable is a discretized version of a continuous variable to reflect how the CE survey program tracks and reports this variable as a matter of practice.}.  The 11th variable, family income, is continuous and considered sensitive due to its high degree of skewness\footnote{Values in Table \ref{tab1} are rounded for confidentiality. Negative family income values reflect investment and business loses.}. As a result, it is synthesized to induce privacy protection. See Table \ref{tab1} for variable details.

\begin{table}[H]
\caption{Variables used in the Consumer Expenditure Survey sample. Data taken from the 2017 Q1 CE. \label{tab1}}
\centering
\begin{tabular}{p{1.2in} p{4.3in} }
\hline
Variable &  Description\\ \hline
Gender & Gender of the reference person; 2 categories \\
Age & Age of the reference person; 5 categories  \\
Education Level & Education level of the reference person; 8 categories  \\
Region & Region of the CU; 4 categories \\
Urban & Urban status of the CU; 2 categories  \\
Marital Status & Marital status of the reference person; 5 categories  \\
Urban Type & Urban area type of the CU; 3 categories  \\
CBSA & 2010 core-based statistical area (CBSA) status; 3 categories \\
Family Size & Size of the CU; 11 categories \\
Earner & Earner status of the reference person; 2 categories  \\
Family Income & Imputed and reported income before tax of the CU; approximate range: (-7K, 1,800K), with 97.5 percentile around \$270K \\ \hline
\end{tabular}
\end{table}

Currently, the CE PUMD releases a topcoded version of the family income values to the public.  While the application of topcoding techniques induces privacy protection by not releasing the exact value of a CU's family income for certain portions of the distribution, topcoding might negatively impact the utility of the microdata by destroying important features of the distribution, especially in the tails. 
There is also an implicit assumption in topcoding that high risk records are concentrated in the right tail of the income distribution, which we show in the sequel to be false.
%
%
%
%
%
%
%
%

The remainder of the paper is organized as follows. Section \ref{marginal} describes the risk-weighted pseudo posterior synthesizer where the weights are constructed from marginal probabilities of identification disclosure. We apply our method to the synthesis of CE family income, and we demonstrate a whack-a-mole problem that arises under marginal downweighting, where downweighting has the unintended affect of increasing the risk on moderate-risk records. We also conduct a simulation study to examine the choice of hyperparameters in our risk-weighted pseudo posterior synthesizer that affects the utility-risk trade-off. Section \ref{pairwise} describes the risk-weighted pseudo posterior synthesizer under a pairwise downweighting framework intended to mitigate the whack-a-mole problem. We demonstrate the proposed pairwise downweighting framework for the synthesis of CE family income and compare its performances of privacy protection and utility preservation to those under the marginal downweighting framework. 
We conclude with a discussion in Section \ref{conclusion}.

\section{Marginal downweighting for risk-weighted synthesis}
\label{marginal}

We denote $y_i$ for the logarithm of the family income for CU $i \in (1,\ldots,n)$, and $\mathbf{X}_i$ for the predictor vector for CU $i$.  The details of our proposed finite mixture synthesizer for the family income variable are included in the Supplementary Material for brevity. We use $\bm{\theta}$ to denote model parameters, and $\bm{\eta}$ for the model hyperparameters.  The estimated $\bm{\theta}$ are subsequently used to produce collections of partially synthetic datasets, $\{\mathbf{Z}^{(l)}, \cdots, \mathbf{Z}^{(L)}\}$, $l \in (1, \ldots, L)$. In each $\mathbf{Z}^{(l)}$, every confidential family income $y_i$ is replaced with a synthesized value, $y_i^{*, (l)}$, drawn from the posterior predictive distribution of the synthesizer. A point estimate and a 95\% confidence interval are calculated from $L$ synthetic datasets combining rules for partially synthetic data.


\subsection{Marginal probability of identification disclosure}
\label{marginal:IR}
We use a marginal probability of identification disclosure, estimated on the \emph{confidential} data, to downweight the likelihood contributions under a risk-weighted pseudo posterior synthesizer.  These probabilities will also be computed and used as risk measures of the resulting \emph{synthetic} data.  We review the details of the pseudo posterior framework in the sequel.  We begin by outlining our method to calculate a marginal probability of identification disclosure for each data record.

The marginal probability of identification disclosure is based on the commonly-used expected match risk measure \citep{ReiterMitra2009}, and adapted to continuous data. We recognize limitations of any risk measures with assumptions of intruder's behavior and knowledge \citep{Hu2019TDP}. 
In practice, owners of confidential data should focus on disclosure risk measures that are most appropriate in their application. 

The released $L$ synthetic datasets, $\mathbf{Z} = (\mathbf{Z}^{(1)}, \cdots, \mathbf{Z}^{(L)})$, are made publicly available by the statistical agency. Suppose an intruder seeks identities of records within each synthetic dataset, $\mathbf{Z}^{(l)}$. In addition to $\mathbf{Z}^{(l)}$, assume the intruder has access to an external file, which contains the following information on each CU $i$: i) a known pattern of the unsynthesized categorical variables, $\mathbf{X}_i = p$, for record $i$. Index $p \in \mathcal{P}$ denotes a pattern that is formed from the intersections of values for \{Gender, Age, Education\} variables; 
ii) the true value of synthesized family income $y_i$; and iii) a name or identity of a CU of interest. An intruder who successfully discovers the identity of a CU is rewarded by accessing the remaining variables in $\mathbf{X}_{i}$ outside the known pattern of variables. With access to such external information, the intruder may attempt to determine the identity of the CU for record $i \in (1,\ldots,n)$ by first performing matches based on the known pattern $p$.   Let $M_{p,i}^{(l)}$ be the collection of CUs sharing the pattern $p$ with CU $i$ for synthetic dataset $\mathbf{Z}^{(l)}$. Let its cardinality, $\lvert M_{p,i}^{(l)}\rvert$, denote the number of CUs in $M_{p,i}^{(l)}$. Note that the intruder will base their interrogation of the records in $M_{p,i}^{(l)}$ on the synthesized family income for those records, \emph{not} their confidential data values (which are unknown to the intruder).

Armed with the knowledge of the true value of family income $y_i$ of CU $i$, the intruder will seek those records whose synthetic values, $y^{\ast,(l)}$, are ``close" to the true value $y_{i}$, where $y^{\ast,(l)}$ denotes the synthesized income value in $\mathbf{Z}^{(l)}$.  Intuitively, if there are few records with synthetic values $y^{\ast,(l)}$ \emph{close} to $y_{i}$, the identification disclosure risk will be higher if the true record is among those records that are close to $y^{\ast,(l)}$.  The identification disclosure risk would be higher because the  true value, $y_{i}$ is not well ``covered."  The intruder is assumed to randomly select a record (since each is otherwise identical in pattern and close to the truth) from only a few records in the case $y_{i}$ is not well covered, elevating the probability to correctly identify record $i$.

To formally define ``close," let $B(y_{i},r)$ denote a ball of radius $r$ around true value $y_{i}$, the family income\footnote{In the case of a univariate variable of interest, $y$, the ball reduces to an interval, though we use the term ``ball" throughout as our framework readily applies to the multivariate case.}.  Let indicator $T_{i}^{(l)} = 1$, if the synthetic data value for record $i$, $y_{i}^{\ast,(l)}$, is among those records, $h \in M_{p,i}^{(l)}$ whose $y_{h}^{\ast,(l)}  \in B(y_{i},r)$.  We set $T_{i}^{(l)} = 1$ if the synthesized value for record $i$ is close to (within a ball of radius $r$) the true value $y_{i}$.   We define the probability of identification for record $i$ as:
\begin{equation}
IR^{(l)}_{i}  = \frac{\mathop{\sum}_{h \in M_{p,i}^{(l)}}\mathbb{I}\left(y_{h}^{\ast,(l)}\notin B(y_{i},r)\right)}{\lvert M_{p,i}^{(l)}\rvert} \times T^{(l)}_{i},
\label{eq:marginalIR}
\end{equation}
where $\mathbb{I}(\cdot)$ is the indicator function.   This identification disclosure risk is constructed as the probability that the synthetic values for records in pattern $p$ are \emph{not} close to the truth.  In other words, if there are relatively few records in the pattern close to the truth, the intruder has a higher probability of guessing the identity they seek.  
%
The choice of $r$ connotes a notion of close that we use to identify (the synthesized value for) a record as isolated using the complement of close.  The value for $r$ is set by the statistical agency as a policy choice.  A more risk-intolerant agency will select a relatively smaller value for $r$, which produces a higher probability that a record is less covered or relatively more isolated.  The higher probability that a record is isolated produces a higher identification risk to the extent that the indicator $T_{i}^{(l)} = 1$, where the synthesized value $y^{\ast, (l)}_{i}$ is close to its true value $y_i$. 
As $r$ shrinks, each record becomes relatively more isolated, though  $T_{i}^{(l)}$ is more likely to flip from $1$ to $0$.  This flipping phenomenon will be more pronounced in relatively well-mixing synthesizers.  

In practice, we recommend setting $r$ through specifying a percentage of $y_{i}$ in order to allow the magnitude of the radius of closeness to adapt to the magnitude of the data value.  Statistical agencies find the setting of $r$ in this way to be more intuitive. The agencies set $r$ based on their notion of closeness for the target variable from their expertise about the class of data. This way of setting $r$ serves as a proxy for guessing the value of $r$ that intruders will use. An intruder draws a radius around a true value known to them and, subsequently, randomly selects a record from those deemed close as the record they seek.   We compute the disclosure risk probabilities under a range of values of $r$ in a simulation study in Section \ref{marginal:sim}, where we demonstrate similar results for risk assessments across a range of neighboring $r$ values.

\begin{figure}[H]
\centering
\includegraphics[width=0.7\textwidth]{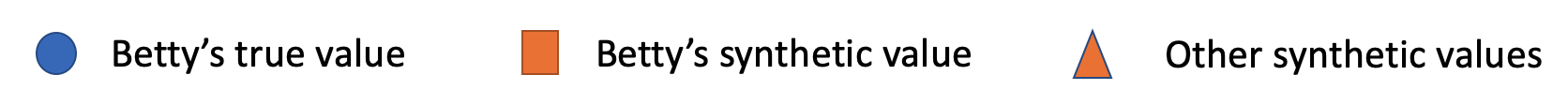}
\end{figure}
\vspace{0mm}
\begin{figure}[H]
    \centering
    \subfloat[$IR_i^{(l)} = \frac{10}{13}$]{ \includegraphics[width=0.25\textwidth]{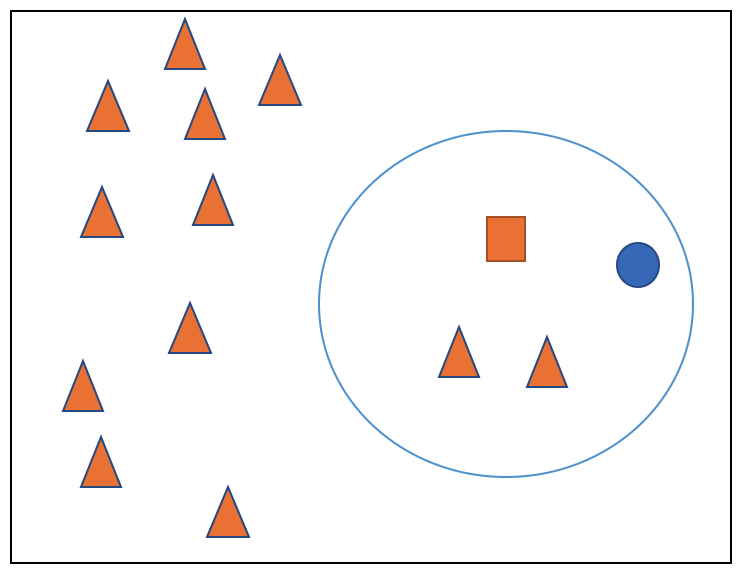}
    \label{fig:case1}}
    \subfloat[$IR_i^{(l)} = \frac{5}{13}$]{ \includegraphics[width=0.25\textwidth]{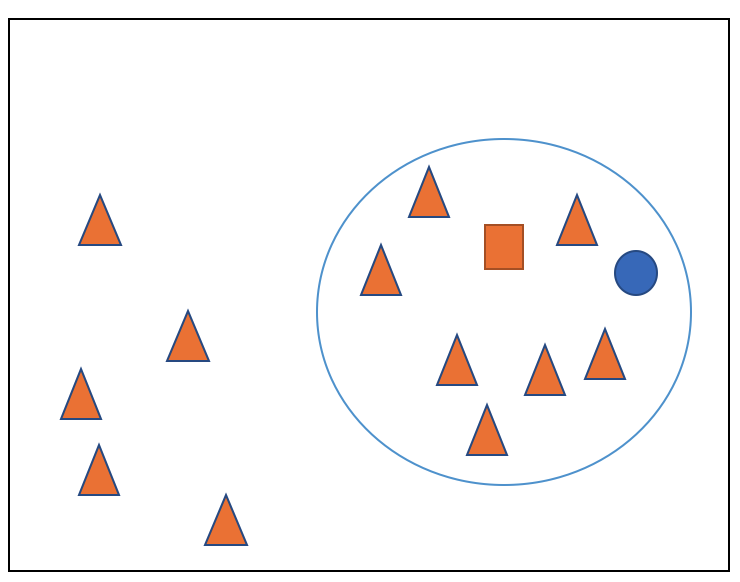}
    \label{fig:case2}}
     \subfloat[$IR_i^{(l)} = 0$]{  \includegraphics[width=0.25\textwidth]{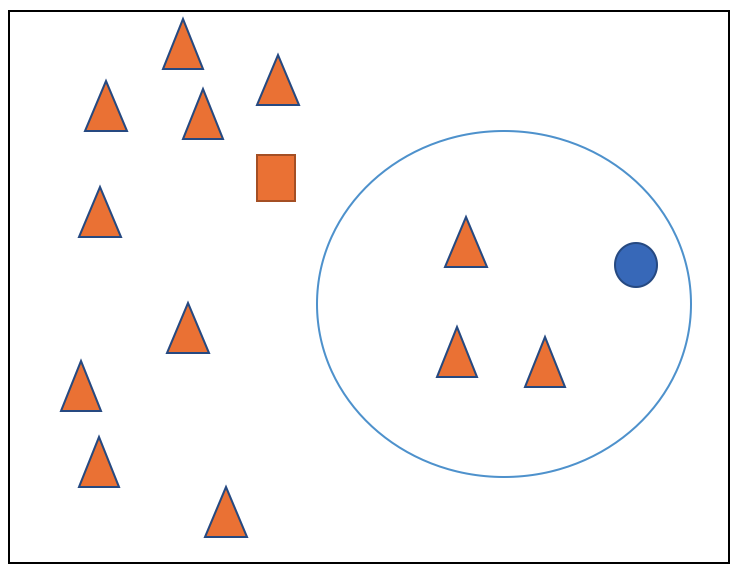}
    \label{fig:case4}}\\
    \caption{Three cases of Betty's marginal identification disclosure probability, $IR_i^{(l)}$: $|M_{p,i}^{(l)}| = 13$ for all three cases, while $\mathop{\sum}_{h \in M_{p,i}^{(l)}}\mathbb{I}\left(y_{h}^{\ast,(l)}\notin B(y_{i},r)\right)$ is 10, 5, and 10 respectively, and $T^{(l)}_{i} = 1$ is 1, 1, and 0, respectively.}
    \label{fig:cases124}
\end{figure}

To illustrate the computations of disclosure probabilities, we provide three toy examples of identifying a person named ``Betty" in the data.  The number of CUs sharing the same pattern as Betty is set to $\lvert M_{p,i}^{(l)}\rvert = 13$. The values of other quantities in Equation (\ref{eq:marginalIR}) are in the figure caption. $r$ is the radius of the ball and assumed the same in all scenarios. Additional toy examples are in the Supplementary Material.

Figure \ref{fig:case1} illustrates the case where few records in the pattern are close to Betty's true record value $y_i$.  This scenario leads to a relatively higher identification disclosure probability because there are fewer options for an intruder to select. Figure \ref{fig:case2} illustrates the case where Betty's true value $y_i$ is well covered by many records in the pattern. This scenario leads to a relatively lower probability of identification disclosure since there are many more equally attractive choices for the intruder. Figure \ref{fig:case4} is similar to the case in Figure \ref{fig:case1}, but now Betty's true value $y_i$ is not close to her synthesized value, which leads to $T_{i}^{(l)} = 0$ and ultimately $IR_i^{(l)} = 0$. 

%
%
%
%
%
%
%
%

We take the average of $IR_i^{(l)}$ across $L$ synthetic datasets and use $IR_i = \frac{1}{L}\sum_{l=1}^{L} IR_i^{(l)}$ as the final record-level identification disclosure risk for CU $i$, which is our identification disclosure risk measure of simulated synthetic data. 



We now turn to the design of pseudo posterior risk-weighted synthesizers, based on the marginal disclosure risk probability of each record in the \emph{confidential} data.

\subsection{Risk-weighted pseudo posterior synthesizer}
\label{marginal:ppsynthesizer}

To create a risk-weighted pseudo posterior synthesizer, statistical agencies calculate the identification disclosure risk of CU $i$ based on the \emph{confidential} data. That is, the calculated record-level identification disclosure probabilities reflect the risks of the confidential data, which will be used in designing a risk-weighted pseudo posterior synthesizer. Calculating disclosure probabilities in the confidential data is a simplified version of Equation~(\ref{eq:marginalIR}), where $IR_i$ reduces $IR_i^c$ below ($c$ denotes confidential data):

\begin{equation}
IR_{i}^c  = \frac{\mathop{\sum}_{h \in M_{p,i}}\mathbb{I}\left(y_{h}\notin B(y_{i},r)\right)}{\lvert M_{p,i}\rvert},
\label{eq:marginalIR_orig}
\end{equation}
where $y_{h}$ denotes a confidential data record in the ball surrounding $y_{i}$ in pattern $p$.

When $IR_i^c$ of CU $i$ is relatively high, the likelihood contribution of CU $i$ will be downweighted more heavily, which in turn strengthens the influence of the prior distribution for CU $i$. Therefore, the record-level weights should be inversely proportional to the identification disclosure risks of the confidential data. We propose the following formulation, which produces record-indexed weights $\in [0,1]$:
\begin{eqnarray}
\alpha_i^m &=& 1 - IR_i^c,
\label{eq:marignalweights}
\end{eqnarray}
where $IR_i^c$ is the identification disclosure risk of CU $i$ in the confidential data. The superscript $m$ denotes that the weight is constructed from the marginal probability of disclosure for record $i$. Suppose CU $i$ expresses a marginal identification risk probability, $IR_i^c = 0.7$, its weight is computed as $\alpha_i^m =1 - 0.7 = 0.3$. 


We formulate the following marginally risk-weighted pseudo-posterior distribution to induce misspecification into our re-estimated synthesizer,
\begin{equation}
\label{eq:pseudo}
p_{\alpha^m}\left(\bm{\theta} \mid \mathbf{y}, \mathbf{X}, \bm{\eta}\right) \propto \left[\mathop{\prod}_{i=1}^{n}p\left(y_{i} \mid \mathbf{X}, \bm{\theta}\right)^{\alpha_{i}^m}\right]p\left(\bm{\theta}\mid \bm{\eta}\right),
\end{equation}
where $\mathbf{X}$ denotes the predictor matrix, $\bm{\theta}$ denotes the model parameters, and $\bm{\eta}$ denotes the model hyperparameters. We then generate a collection of $L$ synthetic datasets, $\bm{Z}$, from this marginally risk-weighted pseudo posterior synthesizer. This procedure surgically distorts the high-risk portions in the data distribution to produce new synthetic datasets with higher privacy protection.

\begin{algorithm}[h]
\setstretch{1.1}
\SetAlgoLined
1. Compute $M_{p,i}$, the set of records in the confidential data sharing the same pattern as record $i$ for $i = 1,\ldots,n$ data records.

2. Cast a ball $B(y_i, r)$ with a radius $r$ around the \emph{true} value of record $i$, $y_{i}$, and count the number of confidential records falling outside the radius $\sum_{h \in M_{p,i}} \mathbb{I}\left(y_h \notin B(y_i, r)\right)$.

3. Compute the record-level risk probability, $IR_i^c$ in the confidential data, as $IR_i^c = \sum_{h \in M_{p,i}} \mathbb{I}\left(y_h \notin B(y_i, r)\right) / \lvert M_{p,i} \rvert$, such that $IR_i^c \in [0, 1]$.

4. Formulate by-record weights, $\bm{\alpha}^{m} = (\alpha_1^{m}, \cdots, \alpha_n^{m})$, $\alpha_i^m =  1 - IR_i^c$.

5. Use $\bm{\alpha}^{m} = \left(\alpha_{1}^{m},\ldots,\alpha_{n}^{m}\right)$ to construct the pseudo likelihood from which the pseudo posterior is estimated. 

6. Estimate parameters from the risk-weighted pseudo posterior,
$
p_{\alpha^m}\left(\bm{\theta} \mid \mathbf{y}, \mathbf{X}, \bm{\eta}\right) \propto \left[\mathop{\prod}_{i=1}^{n}p\left(y_{i} \mid \mathbf{X}, \bm{\theta}\right)^{\alpha_{i}^m}\right]p\left(\bm{\theta}\mid \bm{\eta}\right).
$

7. Draw $L$ synthetic datasets, $\bm{Z} = (\bm{Z}^{(1)},\ldots,\bm{Z}^{(L)})$, from the model posterior predictive distribution.

8. Compute the identification disclosure risks on each synthetic dataset $\bm{Z}^{(l)}$, $l \in (1,\ldots,L)$, using $
IR^{(l)}_{i}  = \frac{\mathop{\sum}_{h \in M_{p,i}^{(l)}}\mathbb{I}\left(y_{h}^{\ast,(l)}\notin B(y_{i},r)\right)}{\lvert M_{p,i}^{(l)}\rvert} \times T^{(l)}_{i},
$
where $y_{h}^{\ast,(l)}$ denotes a synthetic data record in dataset $\bm{Z}^{(l)}$.

9. Take the average of $IR_i^{(l)}$ across $L$ synthetic datasets and use $IR_i = \frac{1}{L}\sum_{l=1}^{L} IR_i^{(l)}$ as the final record-level identification disclosure risk in the synthetic datasets for CU $i$. 
 \caption{Steps to implement the Marginal risk-weighted synthesizer.}
 \label{alg:marginal}
\end{algorithm}
\vspace{1mm}

Our use of weight $\alpha^{m}_i$ applied to the likelihood of observation $y_i$ can be seen as an anti-informative prior, as in $1/\left(p\left(y_{i} \mid \mathbf{X}, \bm{\theta} \right)\right)^{1-\alpha_i^m}$ for CU $i$. The weights ${\bm{\alpha}}^m$ are purposefully designed to partially defeat the likelihood principle to induce misspecification, targeting only the high-risk portion of the distribution. This greater downweighting of high-risk records is what we mean by ``surgical" downweighting as contrasted with an equal downweighting of all records.  Furthermore, the weights are dependent on the confidential data $\bf y$, which the statistical agency holds private and considers \emph{known}. We therefore do not model the weights $\bm{\alpha}^m$ along with $\bf y$ since the by-record risks on which these weights are computed are viewed as \emph{exact}.  
Instead, we use the weights as plug-in to surgically distort high-risk portion of the distribution.

With simulated synthetic data $\bm Z$, we compute the identification disclosure risks, $IR_{i}$ using the general method of Section~\ref{marginal:IR}. We summarize the steps of implementation of the Marginal risk-weighted synthesizer in Algorithm~\ref{alg:marginal}. 

\subsection{Application to synthesis of CE family income $\mathbf{y}$}
\label{marginal:app}

We utilize two synthesizers on the CE sample to synthesize the sensitive family income variable: i) The \emph{unweighted} synthesizer (labeled ``Synthesizer"); ii) The \emph{marginally} risk-weighted pseudo posterior synthesizer of Section \ref{marginal:ppsynthesizer} (labeled ``Marginal"). For comparison, we include results of the topcoded family income (labeled ``Topcoding"). 

For each synthesizer or topcoding procedure, the resulting by-record distribution of the marginal probability of identification distribution in Section \ref{marginal:IR} is computed based on using a radius of closeness of $r = 20\%$ (i.e. the final radius value for record $i$ is $r \times y_i = 0.2 \times y_i$).  The CE program determined that the $20\%$ value meets their determination of income values that are deemed close.  The CE program additionally examined the utility of summary statistics and regression analysis on the synthetic data generated from the risk-weighted pseudo posterior when the weights are computed using $r = 20\%$, and concluded the level of utility was acceptable. 

We assume the intruder has an external file with information on \{Gender, Age, Region\} because these $3$ variables are publicly-released for each of the $n = 6208$ CUs. The intersecting values of these known-to-intruder predictors produces $40$ known patterns, and each pattern contains greater than 1 CU (i.e., no pattern with singletons). 
We generate $L = 20$ synthetic datasets for each synthesizer. We choose $L > 1$ to make sure that the risk profiles to be evaluated and compared are not influenced by any one synthetic dataset, but based on an average across $L = 20$ synthetic datasets. We also choose $L$ sufficiently large to allow the user to make a relatively accurate computation of the between-datasets variance, which is used in the combining rules to capture the model uncertainty. 
Detailed discussion on choices of $L$ is in Section \ref{marginal:sim}.

\subsubsection{Results of risks and utility}
\label{marginal:app:results}

\begin{figure}[H]
\centering
\includegraphics[width=0.5\textwidth]{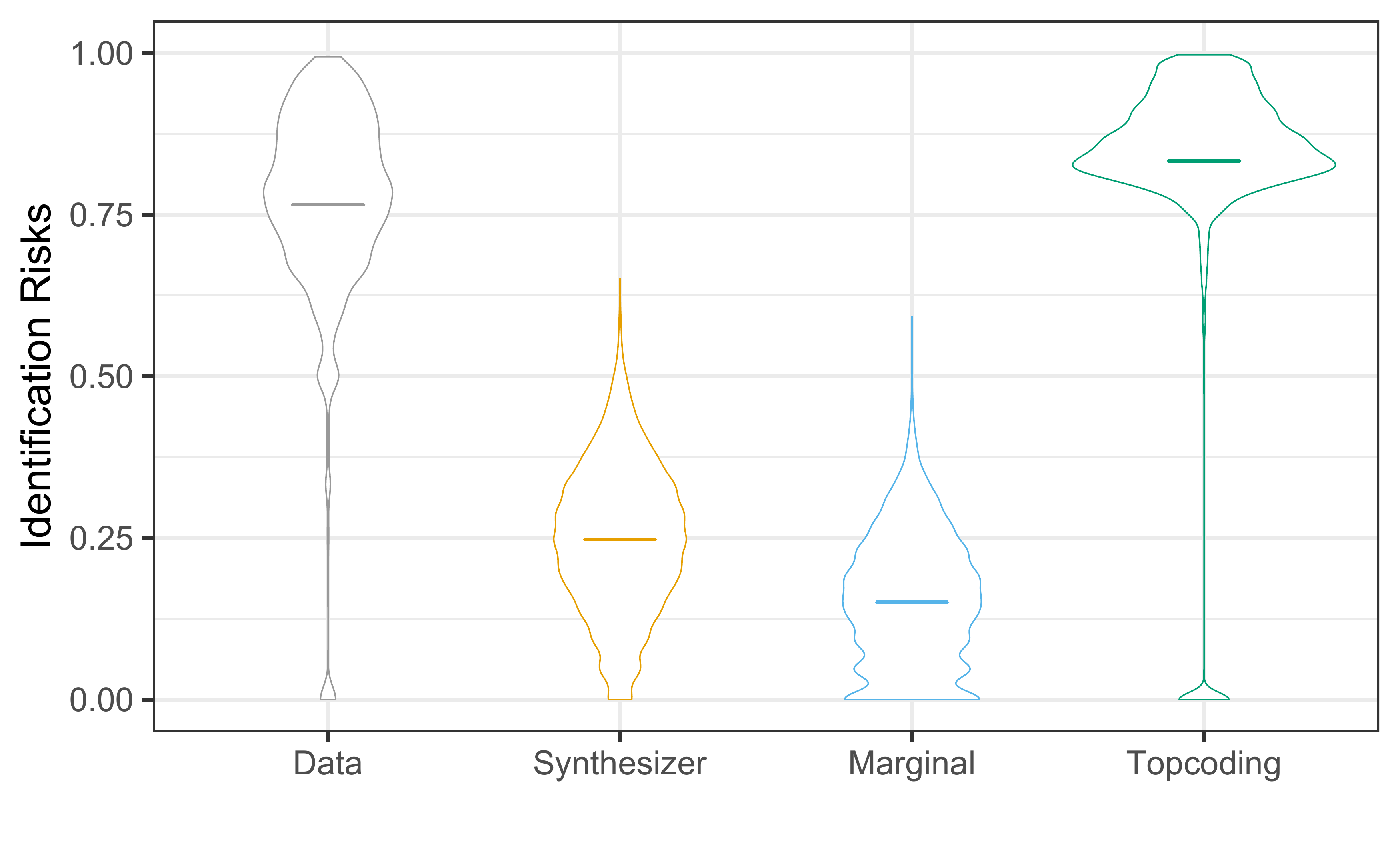}
\caption{Violin plots of the identification risk probability distributions of the CE sample, the unweighted Synthesizer, the Marginal synthesizer, and the Topcoded data. The horizontal bar denotes the mean. 
\label{fig:app:violin1}}
\end{figure}

To compare the risk profiles of the synthetic and topcoded datasets, we use violin (density) plots shown in Figure \ref{fig:app:violin1}. Each violin plot shows the distribution ($\in [0,1]$) of the identification risk probabilities (i.e. $IR_i$'s) for all $n = 6208$ CUs in the CE sample. The higher the identification risk probability of the datasets, the lower the privacy protection associated with their release, and vice versa. The identification risk profile of the confidential data (labeled ``Data") is included for comparison. The columns Synthesizer and Marginal in Table \ref{tab:top10risky} displays the identification risk for the top 10 risky records as measured from confidential data, along with synthetic datasets generated from different procedures. 
Table \ref{tab:top10risky} provides the range of family income values, instead of the actual data value in column ``Value" for confidentiality.

We observe that the Synthesizer produces a significant risk reduction compared to the Data due to prior smoothing of the confidential data distribution. The Marginal further lowers the overall identification disclosure risks as compared to the Synthesizer by surgically downweighting the likelihood contributions of relatively high-risk records. This confirms and demonstrates that our proposed marginal downweighting framework provides higher privacy protection, reducing the peak record-level identification disclosure risks.   Marginal downweighting shrinks or concentrates high-risk records towards main modes of the distribution as we note by nearly $0$ risk values produced for the top 10 risky records by the Marginal in Table \ref{tab:top10risky}.

\begin{table}[H]
\caption{Table of identification risks for top 10 risky records. The highest identification risk case among Synthesizer, Marginal, and Topcoding is bolded for each row. \label{tab:top10risky}}
\centering
\begin{tabular}{ c c c | c c }
\hline
Value & Data & Synthesizer & Marginal & Topcoding \\ \hline
(-10K, 50K) & 0.9944 & 0.0000 & 0.0000 & {\bf 0.9894} \\
(100K, 1000K) & 0.9944 & {\bf 0.0995} & 0.0000 & 0.0000 \\
(-10K, 50K) & 0.9921 & 0.1454 & 0.0000 & {\bf 0.9582} \\
(-10K, 50K) & 0.9915 & 0.0000 & 0.0496 & {\bf 0.9926} \\
(100K, 1000K) & 0.9913 & {\bf 0.0498} & 0.0000 & 0.0000 \\
(-10K, 50K) & 0.9913 & 0.1469 & 0.0000 & {\bf 0.9852} \\
(-10K, 50K) & 0.9913 & 0.0491 & 0.0000 & {\bf 0.9707} \\
(100K, 1000K) & 0.9910 & 0.0000 & 0.0000 & 0.0000 \\
(-10K, 50K) & 0.9910 & 0.1471 & 0.0000 & {\bf 0.9894} \\
(1000K, 2000K) & 0.9910 & 0.0000 & 0.0000 & 0.0000 \\ \hline
\end{tabular}
\end{table}

\begin{table}[H]
\caption{Table of identification risks for top 10 largest records. The highest identification risk case among Synthesizer, Marginal, and Topcoding is bolded for each row. \label{tab:top10size}}
\centering
\begin{tabular}{ c c c | c c }
\hline
Value & Data & Synthesizer & Marginal & Topcoding \\ \hline
(1000K, 2000K) & 0.9910 & 0.0000 & 0.0000 & 0.0000 \\
(1000K, 2000K) & 0.9643 & {\bf 0.0499} & 0.0000 & 0.0000 \\
(1000K, 2000K) & 0.9600 & 0.0000 & 0.0000 & 0.0000 \\
(100K, 1000K) & 0.9820 & 0.0000 & 0.0000 & 0.0000 \\
(100K, 1000K) & 0.9913 & {\bf 0.0498} & 0.0000 & 0.0000 \\
(100K, 1000K) & 0.9565 & 0.0995 & 0.0000 & {\bf 0.9955} \\
(100K, 1000K) & 0.9910 & 0.0000 & 0.0000 & 0.0000 \\
(100K, 1000K) & 0.9872 & {\bf 0.0994} & 0.0000 & 0.0000 \\
(100K, 1000K) & 0.9944 & {\bf 0.0995} & 0.0000 & 0.0000 \\
(100K, 1000K) & 0.9908 & 0.0000 & {\bf 0.0482} & 0.0000 \\ \hline
\end{tabular}
\end{table}

Table~\ref{tab:top10size} presents the probabilities of identification disclosure for the top 10 \emph{largest} family income records. While the intention of topcoding is to provide privacy protection, topcoding only provides such protection on CUs with extremely large family income valueas (as is evident in their long tails and large bulbs around 0 in Figure \ref{fig:app:violin1} and the column Topcoding in Table \ref{tab:top10size}). At the same time, since the majority of the family income is not topcoded (about 6\% of data records are topcoded), topcoding fails to provide \emph{any} privacy protection to most of the CUs that express high risks (as evident in its upper portion in Figure \ref{fig:app:violin1}). Topcoding releases true data values for these high-risk records that are not topcoded, resulting in only a single value near the target income, while yet $T_{i}^{(l)} = 1$ (which is a maximum risk situation).

\begin{table}[H]
\caption{Table of C.I. of mean (top left), median (top right), 90\% quantile (bottom left), and predictor Earner 2 (bottom right) of family income. 
\label{tab:utility_marginal}}
\centering
\begin{tabular}{c | c c | c c}
\hline
 &  estimate & 95\% C.I. & estimate & 95\% C.I.\\ \hline
Data & 72075.60 & [70068.61, 74037.96]  & 50225.15 & [48995.01, 52000.00] \\
Synthesizer & 72377.12 & [69639.31, 75284.18] & 50538.50 &  [48410.30, 52673.39]  \\
Marginal & 76641.95 & [68431.36, 88069.02] & 54229.08 & [51686.27, 57231.27] \\ \hline
Data& 153916.30 & [147582.40, 159603.80]  & -45826.20 & [-49816.29, -41836.11] \\
Synthesizer & 152597.10 & [146203.30, 159957.80] & -46017.29 & [-50239.20, -41795.37] \\
Marginal &  134582.40 & [127342.40, 142549.80] &  -34738.85 & [-46792.90, -22684.81] \\ \hline
\end{tabular}
\end{table}

To compare the utility profiles, we first evaluate analysis-specific utility. In Table \ref{tab:utility_marginal}, we  report  point  estimates  and  95\%  confidence  intervals  of  several summary statistics: the mean, the median, and the 90\% quantile that characterize the distribution of the family income variable, estimated from the collection of $L= 20$ synthetic datasets drawn from the Synthesizer and the Marginal.   All  point  estimates  and  95\%  confidence  intervals  are obtained through bootstrapping.  We also report the regression coefficient  of  Earner  2  (corresponding  to  the  level  that  the  reference  person  is  not an  earner),  in  a  regression  analysis  of  family  income  on  three  predictors, \{Region,Urban, Earner\}.  The point estimates and 95\% confidence intervals for the regression coefficient are obtained by standard combining rules for partially synthetic data. These results indicate that the Synthesizer provides better utility than the Marginal. Second, we evaluate global utility by calculating the empirical CDF metric \citep{Woo2009JPC}, measuring the distance between the confidential dataset and the synthetic datasets. The global utility results also show deteriorated utility of the Marginal, and we include them in the Supplemenatry Material for brevity. 

In summary, our proposed Marginal risk-weighted pseudo posterior synthesizer is able to reduce the disclosure risks further, compared to the unweighted Synthesizer. This risk reduction comes at the price of reduced utility. 

\subsubsection{The whack-a-mole problem}
\label{marginal:app:WAM}


As we have seen, the risk reduction from the Marginal is achieved by shrinking the synthetic data value for each high-risk record to the main modes of the distribution, which in turn reduces its relative isolation from other records (and we note that it is easier for a putative intruder to identify data records with relatively unique values).  The shrinking of a high-risk record towards the main modes, however, may increase the isolation of a relatively moderate-risk record with the result that the identification risk may actually \emph{increase} after estimation under the Marginal as compared to the Synthesizer.  The risk may increase for a moderate-risk record after downweighting because the relatively higher risk record values are shrunk away from that of a moderate-risk record, which leaves the moderate-risk record with reduced coverage.  We term this undesirable phenomenon where the disclosure probability increases for a moderate-risk record after marginal downweighting a ``whack-a-mole." The Marginal unintentionally increases the risk in the synthetic data value of a moderate-risk confidential data record by increasing its relative isolation. It then becomes difficult to control the overall risk profile of the synthetic data under use of the Marginal. 

\begin{figure}[h]
\centering
\includegraphics[width=0.6\textwidth]{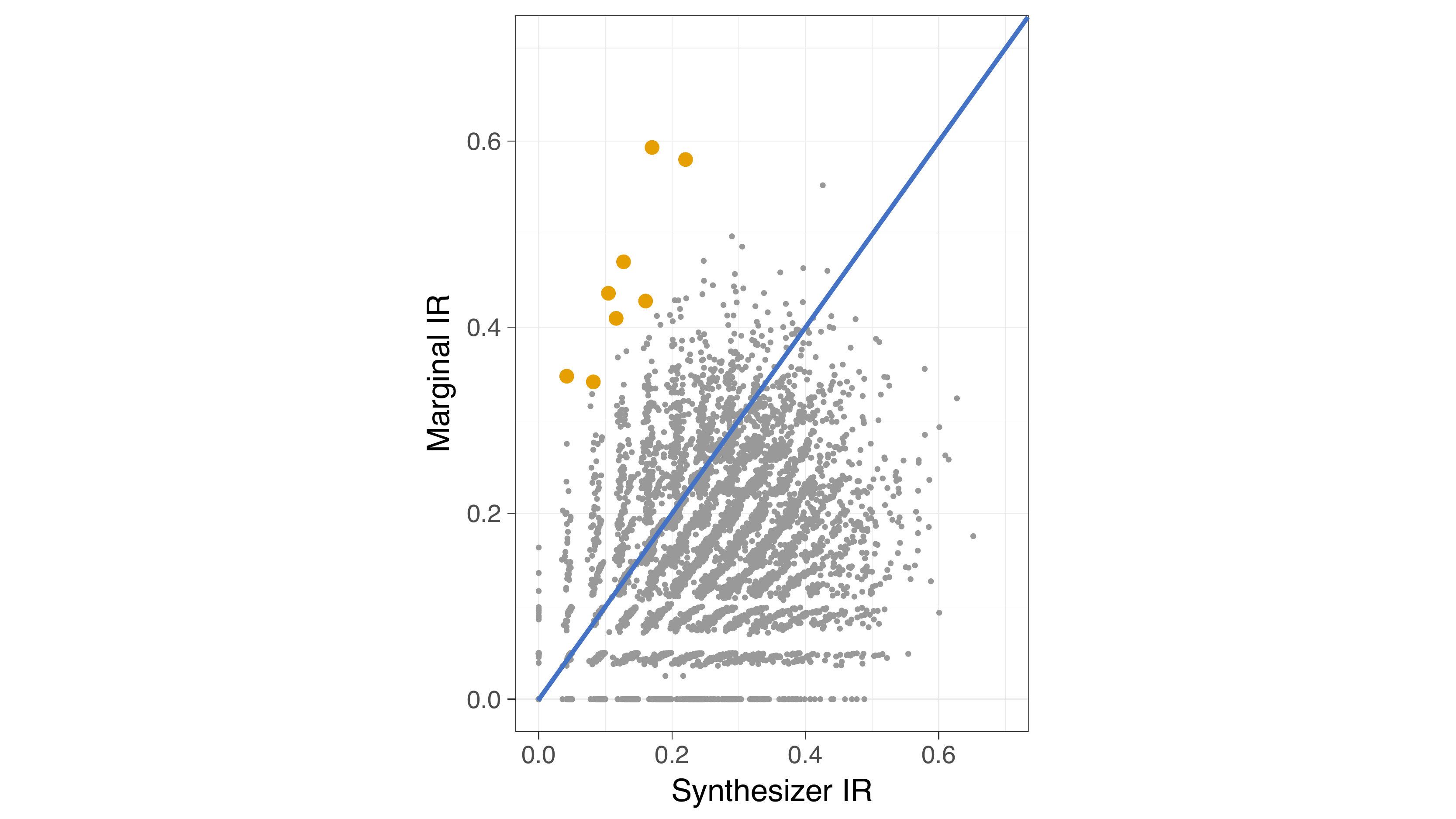}
\caption{Scatterplot of identification risk probabilities of records using the Marginal synthesizer (y-axis) and using the unweighted Synthesizer (x-axis).}
\label{fig:WAM0p25}
\end{figure}

Figure \ref{fig:WAM0p25} highlights records (in yellow with a larger point size) in the CE sample whose marginal probability of identification risk has been increased by 0.25, from the synthetic data drawn under the unweighted Synthesizer to that under the Marginal. This observation suggests that the Marginal sometimes increases the identification risks of records from moderate levels under the Synthesizer to high levels, even though the Marginal successfully provides higher privacy protection for records with high identification risks, and provides an overall lower or downshifted identification risk distribution for the entire dataset as compared to the Synthesizer. We note that the improved overall protection from the Marginal is also evidenced by a majority of records falling under the blue $y=x$ line in Figure \ref{fig:WAM0p25}.

Furthermore, Figure \ref{fig:WAM0p25} shows the Marginal does not control the maximum identification risk well. For example, suppose the statistical agency sets 0.5 as a threshold that no synthetic records should possess identification risk greater than 0.5. As can be seen in the number of records with identification risk exceeding 0.5 on the y-axis, the synthetic data produced by the Marginal does not satisfy such requirement, therefore the set of synthetic data records cannot be released to the public.

We focus on mitigating the whack-a-mole phenomenon to achieve more satisfactory risk profiles of the simulated synthetic data in Section \ref{pairwise}. Our main strategy formulates the weight for each record by constructing a collection of joint, pairwise probabilities of identification risk for that record with the other data records.

We next proceed to demonstrate the generalizability of our risk reduction methods through a simulation study. In addition, we focus on exploring the sensitivity of our methods under a range of values for the radius, $r$, hyperparameter and the number of synthetic released datasets, $L$.

\subsection{Sensitivity analyses of $r$ and $L$ on simulated data}
\label{marginal:sim}

We first demonstrate the generalizability of our risk-weighted pseudo posterior synthesizer by generating simulated count data from a mixture of negative binomial distributions, as a complement to our continuous data application on the CE family income.   We demonstrate an improvement of privacy protection as compared to the unweighted synthesizer that reinforces the same result achieved for the CE dataset.

We proceed to assess the sensitivity of risk and utility profiles to alternative choices of $r$, the radius expressed as a percent of the value for $y_i$. The results reveal that the risk and utility are relatively insensitive to the choice of $r$ within a neighboring range.

Finally, we examine the sensitivity of the estimated variances for statistics generated from the set of $L$ synthetic datasets to the choice of $L$.  We show that for $L \in \{1, 5, 10, 20, 30\}$, we achieve accurate variance computations for statistics that reflect the uncertainty in the synthesizer model estimation with sufficiently large $L$.

\subsubsection{Data generation}
\begin{figure}[H]
  \centering
    \includegraphics[width=0.4\textwidth]{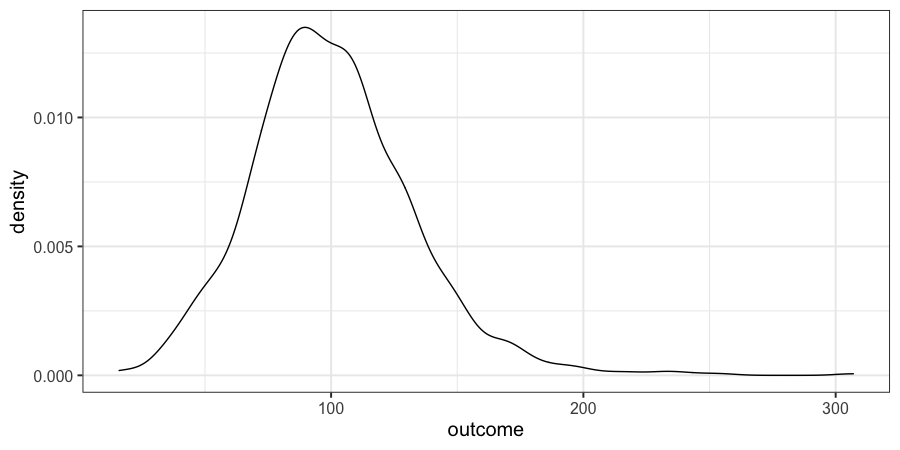}
    \caption{Density plot of the simulated data with skewness.}
    \label{fig:sim_data}
\end{figure}

We generate values from the following 2-component mixture of negative binomial distributions for $n = 1000$ records:
\begin{equation}
    y_{i}\mid \bm{\theta},\bm{\mu},\bm{\phi} \iid \theta_{1}~\mbox{NB}(y_{i}\mid\mu_{1},\phi_{1}) + \theta_{2}~\mbox{NB}(y_{i}\mid\mu_{2},\phi_{2}),
\end{equation}
where $i = 1,\ldots,n$ indexes records and $\mbox{NB}(\cdot)$ denotes a negative binomial kernel parameterized by mean $\mu \in \mathbb{R}^{+}$ and over-dispersion parameter, $\phi \in\mathbb{R}^{+}$. Hyperparmeter $\phi$ controls how much larger is the variance than the mean.  We set mixture component probabilities $\theta_{1} = 0.7$ and $\theta_{2}= 0.3$.  The means are set equally $\mu_{1}=\mu_{2} = 100$, while the first mixture component uses a larger value for $\phi_{1} = 20$, which produces a well-populated mode with relatively less skewness.  The second component, $\phi_{2} = 5$ is set lower to encourage a larger tail.  Taken together, this mixture produces a relatively wide mode with a heavy tail, which we believe is reflective of confidential data distributions\footnote{For example, it may represent a total employment variable for business establishments collected from the Current Employment Statistics (CES) survey administrated by the BLS. More information about the CES is in the Supplementary Materials for further reading.}. 
Figure~\ref{fig:sim_data} displays the distribution for our simulation data.

\subsubsection{Risk results of the marginal risk-weighted synthesizer}
We choose a single negative binomial likelihood with relatively weakly informative priors for the mean and over-dispersion parameter as our synthesizer.  We recall that we may select \emph{any} synthesizer under our risk-weighted pseudo posterior of Equation~(\ref{eq:pseudo}).

\begin{figure}[H]
    \centering
    \subfloat[IR plots with $r \in \{30\%, 25\%, 20\%, 15\%\}.$]{ \includegraphics[width=0.45\textwidth]{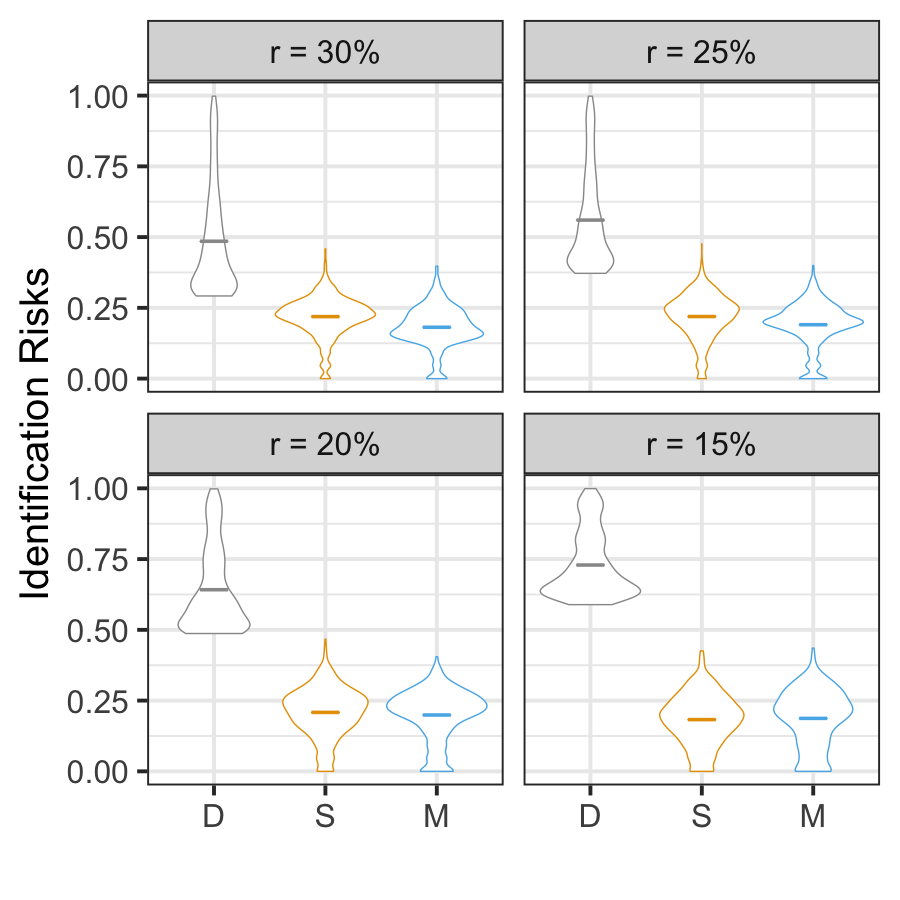}
    \label{fig:compare}}
    \subfloat[Whack-a-model with $r = 15\%.$]{ \includegraphics[width=0.35\textwidth]{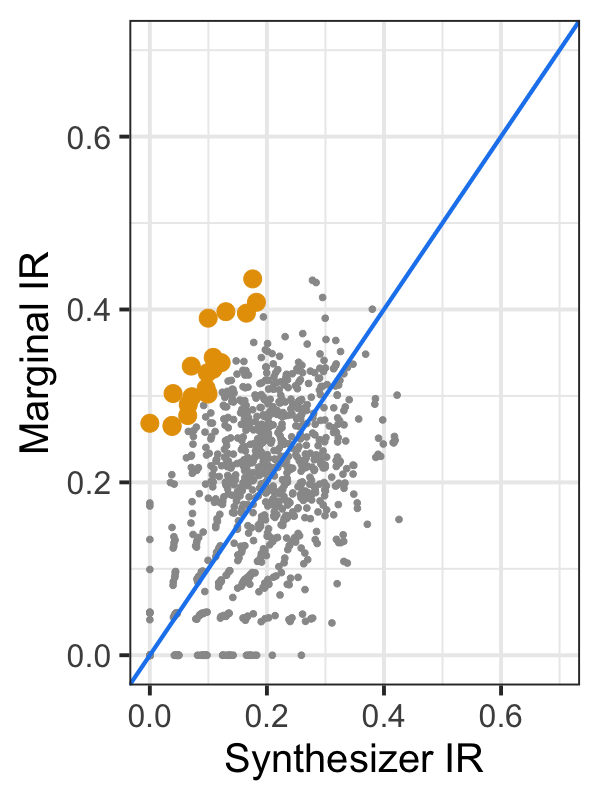}
    \label{fig:sim_wam_marginal}}
    \caption{Risk results of the unweighted Synthesizer and the Marginal synthesizer. The labels in (a) are D for Data, S for Synthesizer, and M for Marginal.}
    \label{fig:sim_risks}
\end{figure}

Each panel in Figure~\ref{fig:compare} presents a comparison of distribution violin plots of the by-record disclosure probabilities that quantifies the risk for each of the confidential data, the unweighted Synthesizer and our Marginal risk-weighted pseudo posterior synthesizer for $r \in \{30\%, 25\%, 20\%, 15\%\}$ with $L = 20$ synthetic datasets.  The figure shows that, for any $r$ value, the unweighted Synthesizer produces a substantial reduction in risk compared to the data, due to smoothing from the use of a negative binomial synthesizer. 
Nevertheless, we see that the Marginal produces a further improvement of risk by shifting the distribution of risk probabilities downward.   

However, for $r = 15\%$, although there is a downward shift from the Synthesizer to the Marginal, we see through the thicker lower ``neck" of the distribution, there is also actually an upshift or increase in the risk probabilities for records with values about the mean.  This upward shift in the riskiness of moderate-risk records reflects the whack-a-mole problem, previously discussed for the CE application.  Figure~\ref{fig:sim_wam_marginal} demonstrates the whack-a-mole problem for the simulated count data under $r = 15\%$ where the enlarged, yellow-colored points highlight records with an increase in risk probabilities from the Synthesizer to the Marginal. 

The whack-a-mole problem is generally expected to become more prominent as $r$ decreases, because outlying records will be computed as relatively higher risk under choices smaller $r$ than larger $r$. Therefore, extremely small $r$ (e.g. $r = 5\%$) could exacerbates the whack-a-mole issue and result in lower risk protection, while larger $r$ (e.g. $r = 70\%$) could result in higher risk protection. Nevertheless, recall that $r$ connotes a notion of closeness for identification. Statistical agencies should set $r$ based on their expertise about the class of data and their level of risk tolerance. Moreover, as we will see next, different values of $r$ demonstrate a utility-risk trade-off, which should be taken into account for statistical agencies when selecting $r$.

\subsubsection{Sensitivity of utility-risk to radius hyperparameter, $r$}
\begin{figure}[H]
  \centering
    \includegraphics[width=0.5\textwidth]{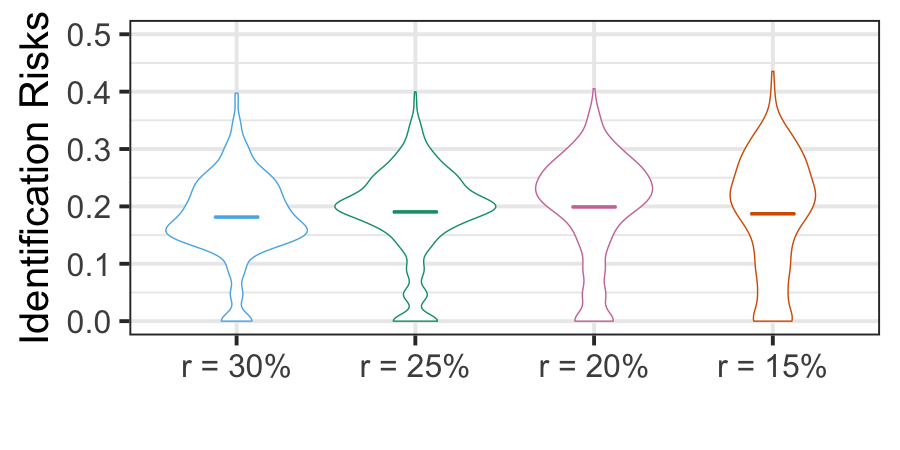}
    \caption{Violin plots of identification risk probabilities for $r \in \{30\%, 25\%, 20\%, 15\%\}$ of the Marginal. The horizontal bar denotes the mean.}
    \label{fig:riskcompare}
\end{figure}

We next explore the sensitivity in risk profiles under varying choices for $r$.  Figure~\ref{fig:riskcompare} presents the distributions of disclosure risk probabilities for decreasing values of $r$ of the Marginal.   Although the average risks are similar, we see that lower values for $r$ tend to produce higher risks as we earlier noted.

\begin{figure}[H]
    \centering
    \subfloat[Mean statistic]{ \includegraphics[width=0.42\textwidth]{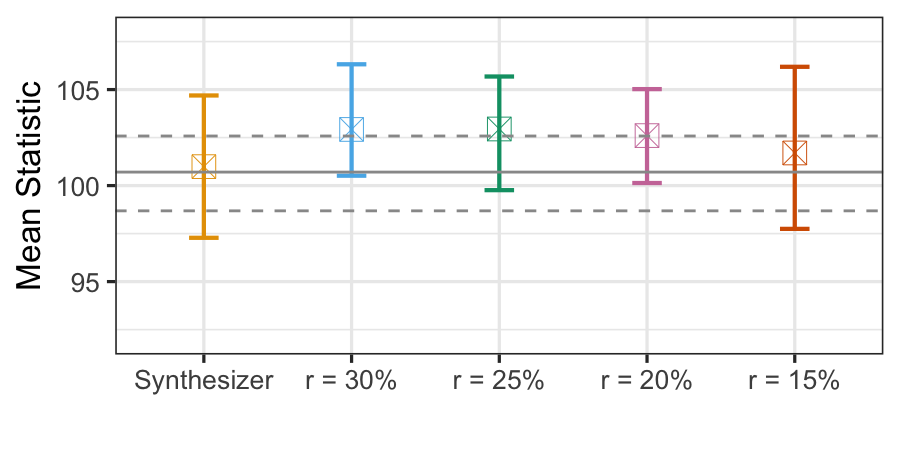}
    \label{fig:NBmixture_utility_mean}}
    \subfloat[Median statistic]{ \includegraphics[width=0.42\textwidth]{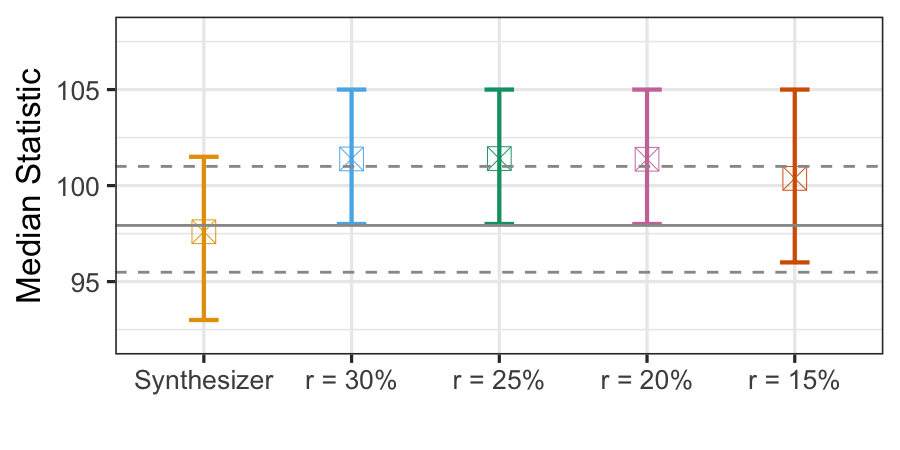}
    \label{fig:NBmixture_utility_median}}\\
    \caption{Utility plots to compare point estimates and 95\% confidence intervals of the Synthesizer and the Marginal for $r \in \{30\%, 25\%, 20\%, 15\%\}$  (cross in a box represents the mean). The horizontal solid line and dashed lines represent the point estimate and the 95\% confidence intervals based on the confidential data, respectively.}
    \label{fig:NBmixture_utility}
\end{figure}

To evaluate analysis-specific utility, Figure~\ref{fig:NBmixture_utility} presents the point estimates and associated $95\%$ bootstrapped confidence intervals of the mean statistic (Figure \ref{fig:NBmixture_utility_mean}) and the median statistic (Figure \ref{fig:NBmixture_utility_median}) of the Synthesizer and the Marginal, where the Marginal is under decreasing values for $r$. 
The solid grey line in each panel presents the confidential data mean value and the grey dashed lines present the associated $95\%$ confidence intervals.  We see that the Synthesizer achieves the highest utility, and the utility of the Marginal slightly improves as $r$ declines, which is coherent with increasing disclosure risk probabilities as $r$ declines, representing a utility-risk trade-off.

\begin{table}[H]
\centering
\caption{Table of empirical CDF utility results for the simulated data.}
\begin{tabular}{l | c c }
\hline
 &  $U_m$ (maximum) & $U_a$ (average squared distance)  \\ \hline
Synthesizer & 0.0451 & 0.0006  \\
$r = 30\%$ &  0.1061 & 0.0040   \\
$r = 25\%$ & 0.1124 & 0.0045  \\
$r = 20\%$ & 0.1152 & 0.0048  \\
 $r = 15\%$ & 0.1088 & 0.0043 \\ \hline
\end{tabular}
\label{tab:ecdf}
\end{table}

To evaluate global utility, we use the empirical CDF metric introduced in Section \ref{marginal:app:results}, which compares the empirical CDFs of the confidential and synthetic datasets.  The first column in Table~\ref{tab:ecdf}, $U_{m}$, presents the maximum difference in CDF values among all records.  Lower $U_m$ indicates higher utility and we see that that $r = 15\%$ produces better utility than $r \in \{20\%,25\%\}$, which matches our analysis-specific utility results.   By contrast $r=30\%$ produces the lowest $U_{m}$ value, though all of the values are relatively close.   The second column presents the $U_{a}$ statistic constructed by computing the squared distance between the CDF values for each record and taking the average.  The same comparative results are obtained as for the $U_{m}$ statistic.

Even though the results demonstrate a utility-risk trade-off across choices for $r$, both the risk profiles and the utility profiles are relatively similar, which indicates relatively little sensitivity in the neighborhood of values of the chosen $r$. 

\subsubsection{Sensitivity in utility profiles to number of synthetic datasets released, $L$}

Finally, we explore the sensitivity in uncertainty quantification to $L$, the number of synthetic datasets released to the public.  The synthesizer utilized by the statistical agency to generate synthetic datasets intended for public release expresses uncertainty in the estimation of model parameters.  Releasing a single (i.e. $L = 1$) synthetic dataset does not include this uncertainty.  Inferential analyses, such as regression, will produce overly optimistic or shortened uncertainty intervals because they exclude the uncertainty in model estimation.  The number of datasets to release, $L$, should be sufficiently large to allow accurate uncertainty quantification. 

Fixing $r = 20\%$, we present mean and median statistic estimates and their $95\%$ confidence intervals under a range of values for $L$ in Table~\ref{tab:L}.  When $L$ is too small, the length of the confidence intervals will be too short.  Therefore, we seek sufficiently large $L$ to allow accurate computations of the confidence intervals.   We see that $L > 10$ does a good job in estimating the confidence intervals well and stably. We also produce our point estimates for any statistic by averaging over the $L$ datasets as a way to stabilize our estimator and reduce dependence on the idiosyncracies of any single dataset.   Table~\ref{tab:L} again reveals that our point estimators stabilize for $L > 10$.   

Therefore, we select $L = 20$ for our analyses to ensure stable computation of the point estimators and accurate estimations of their confidence intervals.

\begin{table}[H]
\centering
\caption{Table of C.I. of mean (left) and median (right) of outcome; results based on the Synthesizer and the Marginal synthesizer with $L \in \{30, 20, 10, 5, 1\}$. The row labels are D for Data, S for Synthesizer, and M for Marginal.}
\begin{tabular}{l | c c c | c c c }
\hline
 &  estimate & 95\% C.I.  & length &  estimate & 95\% C.I. & length \\ \hline
D& {\bf{100.70}} &  [98.68, 102.58]  & {\bf{3.90}} & {\bf{97.93}} & [95.49, 101.00] & {\bf{5.51}} \\ \hline
S ($L = 30$) & {\bf{100.46}} & [96.63, 104.34] & {\bf{7.71}} & {\bf{97.09}} & [93.00, 101.00] & {\bf{8.00}} \\ 
S ($L = 20$)  & {\bf{100.99}} & [97.28, 104.69] & {\bf{7.41}} & {\bf{97.60}} & [93.00, 101.50] & {\bf{8.50}}  \\
S ($L = 10$)  &  {\bf{100.95}} &  [97.07, 104.97] & {\bf{7.90}} &  {\bf{97.36}} & [93.00, 102.00]  & {\bf{9.00}}\\
S ($L = 5$)  & {\bf{99.76}} &  [96.71, 103.33] & {\bf{6.62}} &  {\bf{96.59}} & [93.00, 101.00] & {\bf{8.00}} \\
S ($L = 1$)  & {\bf{102.05}} &  [99.97, 104.04] & {\bf{4.07}} &  {\bf{99.82}} & [97.00, 102.00] & {\bf{5.00}} \\ \hline
M ($L = 30$) & {\bf{102.37}} & [99.49, 104.87] & {\bf{5.38}}  & {\bf{101.05}} & [98.00, 103.50] & {\bf{5.50}}\\
M ($L = 20$) & {\bf{102.60}} & [100.13, 105.02] & {\bf{4.89}} & {\bf{101.37}} & [98.00, 105.00] & {\bf{7.00}} \\ 
M ($L = 10$) &  {\bf{102.93}} & [99.92, 105.34] & {\bf{5.42}} &  {\bf{101.63}} &  [98.00, 105.00] & {\bf{7.00}} \\ 
M ($L = 5$) &  {\bf{102.26}} & [99.69, 104.64] & {\bf{4.95}} &  {\bf{101.00}} &  [98.00, 104.00] & {\bf{6.00}} \\ 
M ($L = 1$) &  {\bf{103.78}} & [102.34, 105.12] & {\bf{2.78}} &  {\bf{103.15}} & [101.00, 105.00] & {\bf{4.00}} \\ \hline
\end{tabular}
\label{tab:L}
\end{table}

\section{Pairwise downweighting for risk-weighted synthesizers}
\label{pairwise}

Our CE data application and simulation study results demonstrate that the Marginal risk-weighted pseudo posterior synthesizer achieves reduced disclosure risks compared to the unweighted Synthesizer. The Marginal performs surgical downweighting of high-risk portions of the data to provide overall higher privacy protection. This risk reduction comes at the price of reduced utility, and it could also have the unintended whack-a-mole effect of exposing moderate-risk records under the Synthesizer to higher risk under the Marginal. We now describe our mitigating approach to achieving more satisfactory risk profiles of the risk-weighted synthesizers.

\subsection{Risk-weighted pseudo posterior  synthesizer}
\label{pairwise:ppsynthesizer}

Similar to creating the Marginal in Section \ref{marginal:ppsynthesizer}, statistical agencies calculate the identification disclosure risk of each record based on the \emph{confidential} data, which will be used to create a pseudo posterior risk-weighted synthesizer. 

Our mitigating approach constructs a pairwise identification risk probability for each pair of CUs $(i,j)$ that are in the \emph{same} known pattern of un-synthesized variables.  Let $M_{p,(i,j)}$ index the collection of CUs in the confidential data sharing the same pattern $p$, with CUs $i$ and $j$ in the confidential dataset. As before, $|M_{p,(i,j)}|$ denotes the number of CUs in the $M_{p,(i,j)}$ collection. Moreover, $B(y_{i}, r)$ casts a ball of radius $r$ around the true family income of CU $i$, $y_i$; similarly, $B(y_{j}, r)$ for CU $j$.  We next measure the probability of the event that the family income in the confidential data for each CU, $h \in M_{p,(i,j)}$, lies in the intersection defined by $y_h \in B(y_{i}, r)$ \emph{and} $y_h \in B(y_{j}, r)$, jointly.  These intersections are used to construct a joint identification risk probability in the \emph{confidential} data, $IR_{i,j}^c$ for the pair of CUs $(i,j)$ as:
\begin{equation}
IR_{i, j}^c = \frac{\mathop{\sum}_{h \in M_{p,(i,j)}}\mathbb{I}\left(y_h \notin B(y_{i},r) \cap  y_h \notin B(y_{j},r)\right)}{\lvert M_{p,(i,j)}\rvert}.
\label{eq:pairwiseIR}
\end{equation}
For pairs of CUs $(i,j)$ assigned to \emph{different} known patterns, the joint identification risk probability is set to $0$ (i.e. $IR_{i,j}^c = 0$).  
Next, we use these joint identification risk probabilities to formulate dependent, by-record probability-based weights.

To construct pairwise weight $\alpha_i^{pw}$ for CU $i$, we use the collection of pairwise identification risk probabilities, $IR_{i,j}^c$ of the confidential data.  The superscript, $pw$, denotes the the weight for each record $i$ is constructed by averaging over the pairwise probabilities for all pairs, $j$, containing $i$ in its pattern.  We construct $\alpha_i^{pw}$ from the collection of pairwise weights, $\alpha_{i,j}$, for each $j \in M_{p,(i,j)}$, defined as:
\begin{equation}
\alpha_{i,j} = 1 - IR_{i,j}^c.
\label{eq:pairwiseweights1}
\end{equation}  
This definition constructs the pairwise weight $\alpha_{i,j}$ to be inversely proportional to the pairwise identification risk probability $IR_{i,j}^c$: higher $IR_{i,j}^c$ results in lower $\alpha_{i,j}$, and vice versa. Furthermore, $\alpha_{i,j} \in [0, 1]$.

We proceed to construct a normalized weight, $\tilde{\alpha}_i \in [0,1]$ for CU $i$, by summing over all $\alpha_{i,j}$ for $j \neq i$, and dividing by $|M_{p,i}| - 1$ to account for the $|M_{p,i}| - 1$ times that CU $i$ appears in the combination of pairs in pattern $p$:
\begin{equation}
\tilde{\alpha}_{i} = \frac{\sum_{j \neq i, j \in M_{p,i}}\alpha_{i,j}}{|M_{p,i}| - 1}.
\label{eq:pairwiseweights2}
\end{equation}

The normalized weight, $\tilde{\alpha}_i \in [0, 1]$, reflects the amount of downweighting needed for CU $i$, based on the sum over all pairwise identification risk probabilities associated with CU $i$, $\{IR_{i,j}^c, j \neq i, j \in M_{p,i}\}$. To see the inverse proportionality between $\tilde{\alpha}_i$ and $\{IR_{i,j}^c, j \neq i, j \in M_{p,i}\}$ more clearly, we can rewrite Equation (\ref{eq:pairwiseweights2}) in terms of $IR_{i,j}^c$:
\begin{eqnarray}
\tilde{\alpha}_{i} &=& \frac{\sum_{j \neq i, j \in M_{p,i}}\alpha_{i,j}}{|M_{p,i}| - 1} =  \frac{\sum_{j \neq i, j \in M_{p,i}}(1 - IR_{i,j}^c)}{|M_{p,i}| - 1} \nonumber \\
	     		&=&  \frac{|M_{p, i}| - 1 - \sum_{j \neq i, j \in M_{p,i}}IR_{i,j}^c}{|M_{p,i}| - 1} = 1 - \frac{\sum_{j \neq i, j \in M_{p, i}}IR_{i,j}^c}{|M_{p,i}| - 1}.
\label{eq:pairwiseweights3}
\end{eqnarray}

Equation (\ref{eq:pairwiseweights3}) shows that when the sum of pairwise identification risk probabilities associated with CU $i$ is high, the normalized weight $\tilde{\alpha}_i$ will be low and close to 0. Such inverse proportionality is desired, because we want to produce more downweighting for CUs with high risks for further privacy protection.  

We recall that the whack-a-mole phenomenon in the marginal downweighting framework arises as distribution mass is shifted in the synthetic data from the confidential data, due to shrinking of family income values for isolated high-risk records towards the main modes of the distribution.  This shrinking of values for high-risk records may in turn reduce the number of records whose values are close to or cover a moderate-risk record. 

By contrast, under the pairwise framework, the set of $(\tilde{\alpha}_{i})$ within each pattern are constructed as \emph{dependent}.  These probability-based weights are therefore formulated to reduce the degree of shrinking of each high-risk record and leave moderate-risk records more covered in the synthetic data, such that the risks of these records increase less than those in the synthetic data under the marginal downweighting framework.  We demonstrate in Section \ref{pairwise:app} that the pairwise downweighting framework induces a compression in the distribution of by-record identification risk probabilities in the resulting synthetic data, due to the dependence among the $(\tilde{\alpha}_{i})$.  This compression helps reduce the whack-a-mole phenomenon.

The final pairwise probability-based weight, $\alpha_i^{pw}$, for CU $i$, is defined as $\alpha_i^{pw} = \tilde{\alpha}_i$. Our pairwise risk-weighted pseudo posterior synthesizer has the same form as specified in Section \ref{marginal:ppsynthesizer} for the marginal risk-weighted pseudo posterior synthesizer; namely,
\begin{equation}
p_{\alpha^{pw}}\left(\bm{\theta} \mid \mathbf{y}, \mathbf{X}, \bm{\eta}\right) \propto \left[\mathop{\prod}_{i=1}^{n}p\left(y_{i} \mid \mathbf{X}, \bm{\theta}\right)^{\alpha_{i}^{pw}}\right]p\left(\bm{\theta}\mid \bm{\eta}\right),
\label{eq:pairwisesynthesizer}
\end{equation}
where with predictor matrix $\mathbf{X}$, model parameters $\bm{\theta}$ , and model hyperparameters $\bm{\eta}$. We then generate a collection of $L$ synthetic datasets, $\mathbf{Z}$, from this pairwise risk-weighted pseudo posterior synthesizer.

In summary, we compute the pairwise identification disclosure risk probabilities on the confidential data, from which we form record-level weights. The weights are used in a Pairwise risk-weighted pseudo posterior synthesizer, and the disclosure risks of the resulting synthetic data are evaluated under the marginal identification disclosure probability defined in Section \ref{marginal:IR}. We include the algorithm to implement the Pairwise risk-weighted synthesizer in the Supplementary Material for further reading. 

\subsection{Application to synthesis of CE family income}
\label{pairwise:app}

We now apply the Pairwise risk-weighted pseudo posterior synthesizer (labeled ``Pairwise") to the CE sample. The Synthesizer and the Marginal are from Section \ref{marginal:app}.
As before, the resulting by-record distribution of identification risks based on the marginal probability of identification risk in Section \ref{marginal:IR} are evaluated for synthetic datasets drawn under each of three synthesizers with $r = 20\%$ with the same available external file with information on \{Gender, Age, Region\}.  
We next evaluate and compare the profiles of identification risks and utility preservation for all synthesizers.

\subsubsection{Identification disclosure risks}

\begin{figure}[H]
\centering
\includegraphics[width=0.45\textwidth]{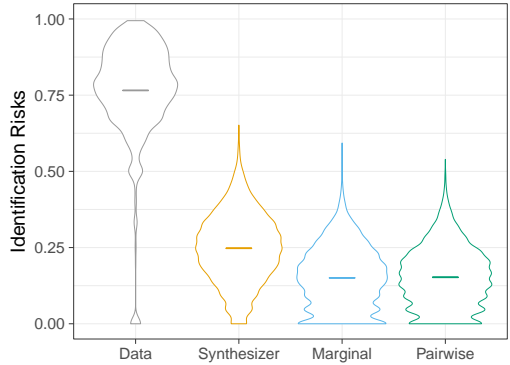}
\caption{Violin plots of the identification risk probability distributions of the CE sample, the unweighted Synthesizer, the Marginal synthesizer, and the Pairwise synthesizer. The horizontal bar denotes the mean.}
\label{fig:IRc1p0}
\end{figure}

Our identification risk evaluation follows the approaches in Section \ref{marginal:app:results} and presented in Figure \ref{fig:IRc1p0}. Recall that the higher the identification risk probability, the lower the privacy protection, and vice versa. 

Between the two risk-weighted pseudo posterior synthesizers, the Pairwise provides a shorter tail, as well as a more concentrated identification risk distribution, compared to the Marginal. With similar average identification risk probabilities (the horizontal bars), the Pairwise has an inter quartile range (IQR) of 0.1385, compared to 0.1534 of Marginal, highlighting that the Pairwise induces a compression in the by-record identification risks computed on the synthetic data under the pairwise downweighting framework. All to say, the Pairwise produces a relatively lower, more compressed identification risk distribution than that of the Marginal in a fashion that offers more control to the BLS.  Constructing the by-record, pairwise probability-based weights, $(\alpha_{i}^{pw})$ to be dependent ties the shrinking of records together in a fashion that reduces the loss of coverage for moderate-risk records in the resulting synthetic data as compared to the Marginal.  We also note that the overall mean identification risk across the records is essentially the same for both the Marginal and the Pairwise as shown in the horizontal line within in each violin plot, indicating that both distributions are centered similarly.  It is the difference in the relative concentration of their masses and the maximum identification risk values that differentiate them.

Furthermore, we note that there is a substantial downward shift in the identification risk distributions for the Marginal and the Pairwise, as compared to the Synthesizer.  This shift may be seen by focusing on the bottom portion of the distributions where there is much more distribution mass for identification risks (measured as marginal probabilities of identification risk) $ < 0.25$.

\begin{figure}
    \centering
    \subfloat[Marginal]{   \includegraphics[width=0.5\textwidth]{WAM_marginal_0p25_line}
      \label{fig:WAM0p25_marginal_dub}}
    \subfloat[Pairwise]{    \includegraphics[width=0.5\textwidth]{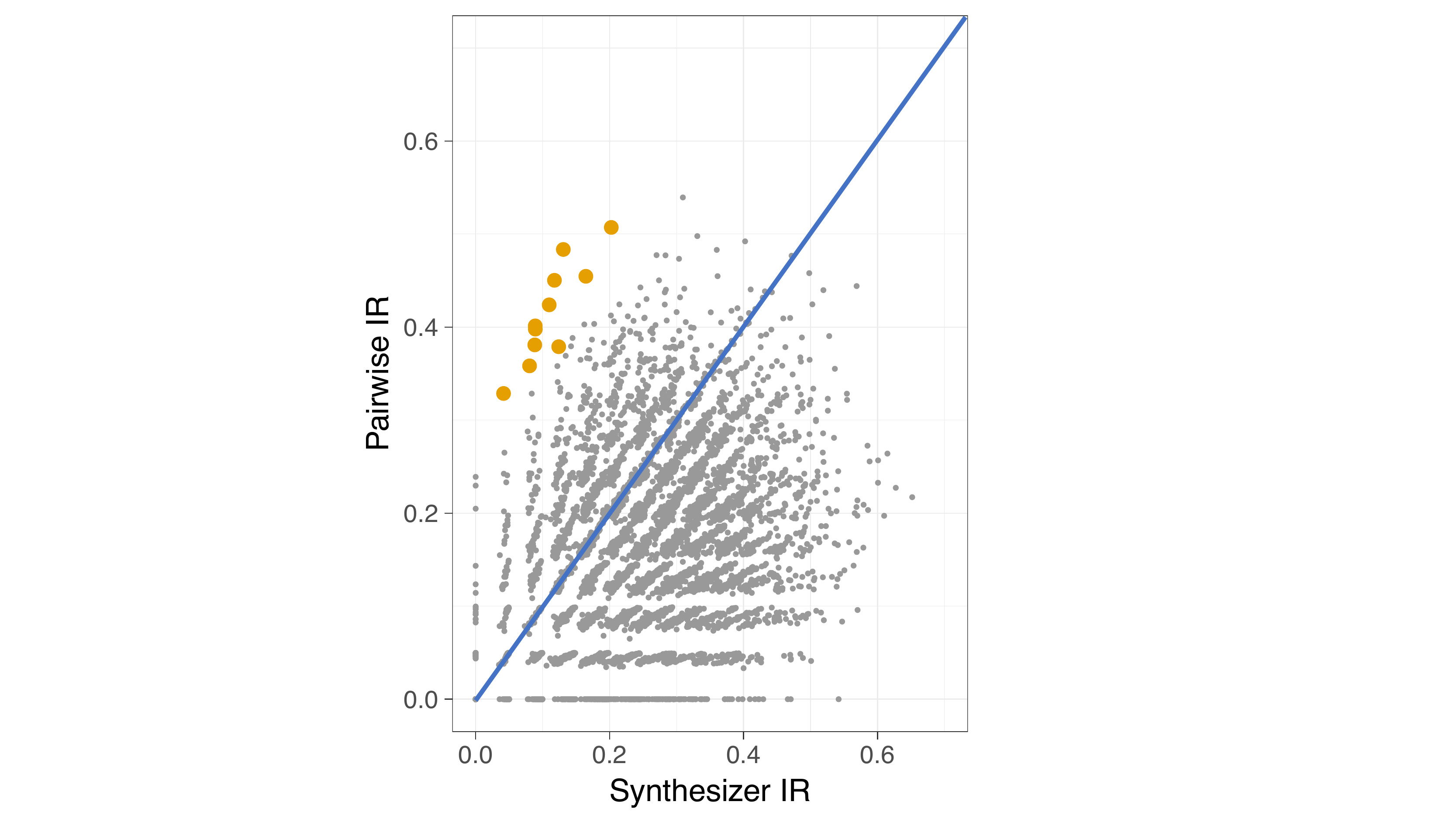}
    \label{fig:WAM0p25_pairwise}}
    \caption{Scatterplot of identification risk of records using the Marginal / Pairwise synthesizer (y-axis) and using the unweighted Synthesizer (x-axis).}
    \label{fig:data_WAM}
\end{figure}

It bears noting that the Pairwise has \emph{not} fully resolved the whack-a-mole phenomenon, though it has notably lessened it. Figure \ref{fig:WAM0p25_pairwise} depicts a scatterplot highlighting the whack-a-mole phenomenon in the Pairwise. Compared to that in the Marginal in Figure \ref{fig:WAM0p25_marginal_dub}, the whack-a-mole phenomenon in the Pairwise is less severe, as can be seen by the overall smaller values on the y-axis of the highlighted CUs (whose identification risk has increased by 0.25 from the Synthesizer to the Pairwise). We also observe in Figure \ref{fig:WAM0p25_pairwise} that there are fewer CUs with higher than 0.5 identification risk in the Pairwise than in the Marginal, a feature that also reduces the tail length in the identification risk distribution violin plot shown in Figure \ref{fig:IRc1p0}. 

We also examined three-way and four-way identification risk probabilities to assess whether further improvement in the whack-a-mole phenomenon was observed, but discovered little improvement at the price of a less scalable computation.

\subsubsection{Utility}

\begin{figure}[H]
    \centering
    \subfloat[Mean statistic]{ \includegraphics[width=0.42\textwidth]{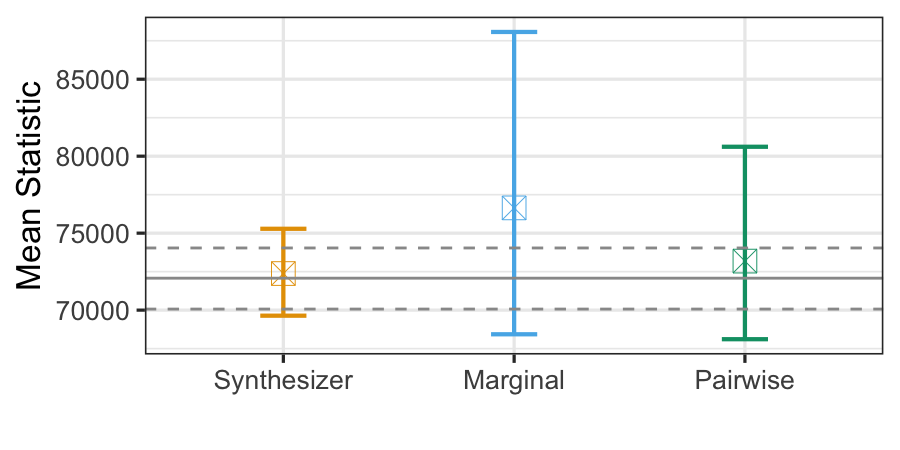}
    \label{fig:data_utility_mean}}
    \subfloat[Median statistic]{ \includegraphics[width=0.42\textwidth]{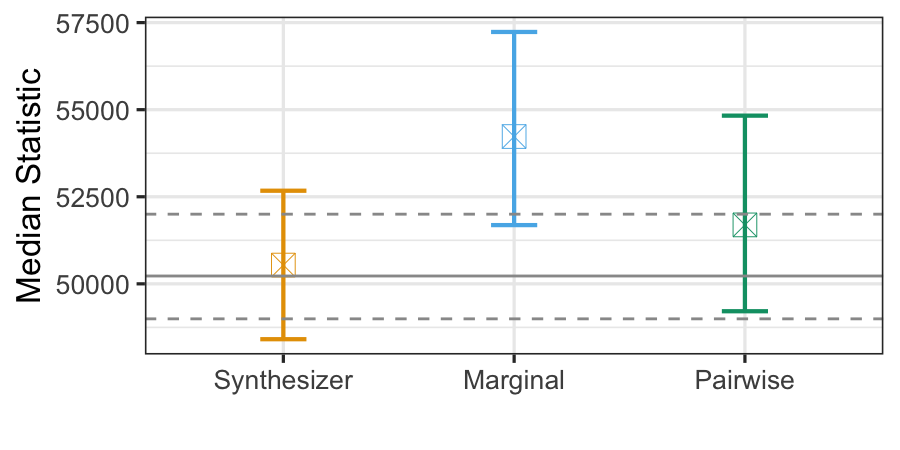}
    \label{fig:data_utility_median}}\\
     \subfloat[90\% quantile statistic]{  \includegraphics[width=0.42\textwidth]{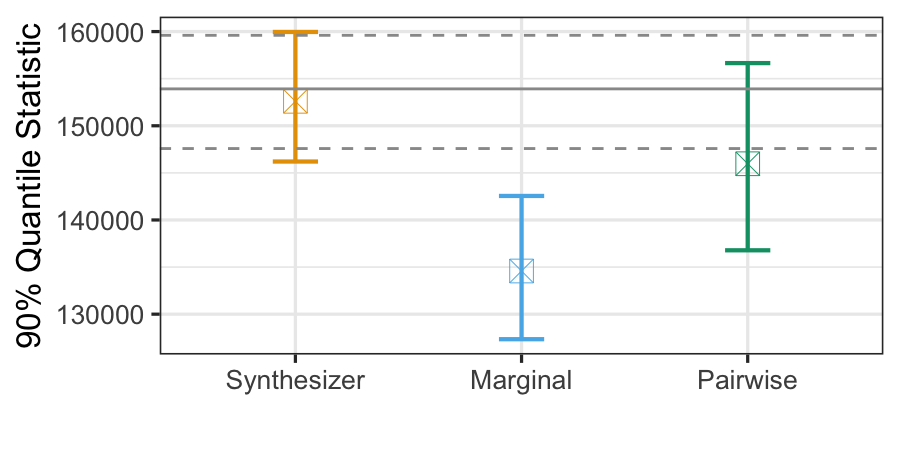}
    \label{fig:data_utility_q90}}
     \subfloat[Earner 2 regression coefficient]{  \includegraphics[width=0.42\textwidth]{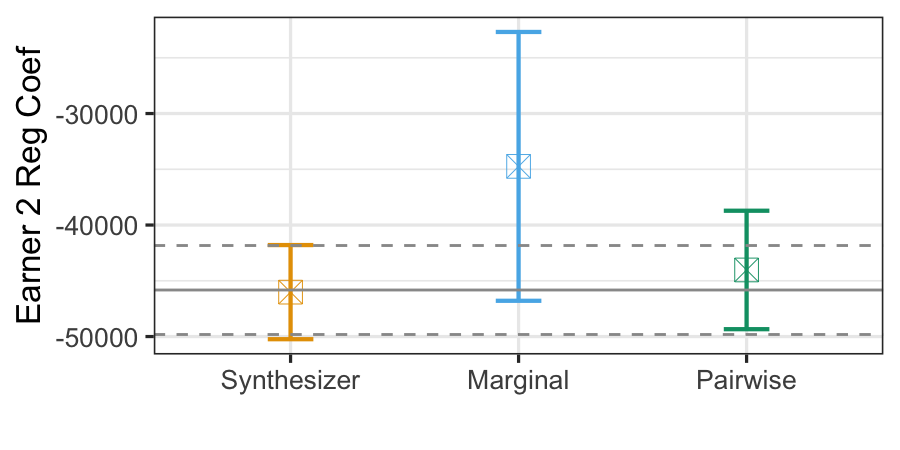}
    \label{fig:data_utility_earner2}}\\
    \caption{Utility plots, comparing point estimates and 95\% confidence interval coverage of Synthesizer, Marginal, and Pairwise (cross in a box represents the mean). The horizontal solid line represents the point estimate, and the horizontal dashed lines represent the 95\% confidence intervals based on the confidential data.}
    \label{fig:data_utility}
\end{figure}

Our analysis-specific utility evaluation follows the approaches in Section \ref{marginal:app:results}, and we present the results in Figure \ref{fig:data_utility}. 
As we know, the Synthesizer maintains a high level preservation of utility at the price of too high a risk profile.

Additional risk reduction is offered by the two risk-weighted pseudo posterior synthesizers, but at the cost of some loss of utility in their synthetic datasets.  The utility results in Figure \ref{fig:data_utility} show that such privacy protection comes at a probably unacceptable utility reduction in the Marginal because the resulted 95\% confidence intervals are either too wide or have little or none overlap with those from the Data (horizontal dashed lines), not to mention that some do not contain the point estimate from the Data (horizontal solid lines).

The Pairwise, nevertheless, also expresses some utility reduction as compared to the Synthesizer, though the reduction is relatively minor such that inference is unchanged. 
The Pairwise achieves higher level of utility preservation than the Marginal because it shrinks the tail or extreme observations \emph{less} to achieve the same risk level. By pairing records when computing risks, the Pairwise covers each risky record with other records whose values are nearby.  This extra coverage leaves these records with less risk as compared to the Marginal, which shrinks records one-at-a-time, such that high-risk records may be over-shrunk towards a mode of the distribution of income values. Based on these results, we recommend the Pairwise to the BLS as a solution that offers further privacy protection while maintaining a reasonably high utility.

\begin{table}[H]
\centering
\caption{Table of empirical CDF utility results.}
\begin{tabular}{l | c c}
\hline
 &  $U_m$ (maximum) & $U_a$ (average squared distance)  \\ \hline
 Synthesizer & 0.0151 & 3.6e-05  \\
Marginal &  0.0755 & 0.0018  \\
Pairwise & 0.0356 & 0.0004 \\ \hline
\end{tabular}
\label{tab:pwecdf}
\end{table}

Finally, we also include the global utility metric of empirical CDFs, that compares the empirical distributions of the real and synthetic datasets in Table~\ref{tab:pwecdf}.  The maximum and average distance statistics show the Pairwise to be twice as good as the Marginal in estimating the confidential data distribution.

%

Results of the Pairwise on simulated 2-component mixture of negative binomial data are similar to the CE application, which are included in the Supplementary Material for brevity. We also propose two additional weight adjustments to improve the level of utility preservation with acceptable loss of privacy protection. These strategies could allow the BLS and other statistical agencies to further tune the risk-weighted pseudo posterior synthesizers to achieve their desired utility-risk trade-off. Details are included in the Supplementary Material for further reading. 

\section{Conclusion}
\label{conclusion}

We propose a general framework for statistical agencies to achieve desired utility-risk trade-off balance when disseminating microdata through synthetic data. Starting with a synthesizer with high utility but unacceptable level of identification risks, statistical agencies can proceed to create risk-weighted pseudo posterior synthesizers to provide a higher degree of privacy protection. Our proposed risk-weighted pseudo posterior synthesizers formulate a record-indexed weight $\in [0, 1]$, which is inversely proportional to the record-level identification risk probability. The likelihood contribution of each record is exponentiated with the record-indexed weight and used to downweight the likelihood contributions of records with high identification risks, providing higher privacy protection.

The agencies may begin with marginal identification risk probabilities applied to \emph{any} unweighted synthesizer 
to improve privacy protection as compared to the unweighted synthesizer.  The marginal identification risk probabilities are formulated to treat the probabilities of disclosure risk for the records to be independent from each other. When risk-weighted pseudo posterior synthesizers based on the marginal identification risk probabilities do not provide sufficient privacy protection, especially in the case where moderate-risk records are exposed to less privacy protection due to the whack-a-mole phenomenon, the statistical agencies may utilize pairwise identification risk probabilities to construct the by-record weights. The pairwise identification risk probabilities tie pairs of records together and induce dependence among the by-record weights, which offers an overall higher privacy protection and mitigates the whack-a-mole phenomenon. 

Our applications to the CE sample and simulation study show that the pairwise downweighting framework creates risk-weighted pseudo posterior synthesizers with better control of the identification risks and little loss of utility. 
These features provide general guidelines for statistical agencies to design risk-weighted pseudo posterior synthesizers to work towards disseminating synthetic data that achieves their desired utility-risk trade-off. We use a flexible finite mixture synthesizer for our CE application and a negative binomial synthesizer for our simulated data, and illustrate that our proposed downweighting frameworks, marginal or pairwise, are applicable to different synthesizers. 


We believe future research is needed to extend our methods to partial multivariate synthesis or full synthesis, and investigate how the choice of $r$ affects the utility-risk trade-off. Results from such research will allow statistical agencies to set informed values of $r$. We note that in the case of multivariate synthesis, the ball of radius $r$ becomes multi-dimension.

Our source code is available at: \url{https://github.com/monika76five/Risk_Efficient_Bayesian_Synthesis}


\subsection*{Acknowledgements}

This research is supported by ASA/NSF/BLS Senior Research Fellow Program.

\bibliography{CEbib}

\begin{thebibliography}{21}
\expandafter\ifx\csname natexlab\endcsname\relax\def\natexlab#1{#1}\fi
\expandafter\ifx\csname url\endcsname\relax
  \def\url#1{\texttt{#1}}\fi
\expandafter\ifx\csname urlprefix\endcsname\relax\def\urlprefix{URL: }\fi

\bibitem[{An and Little(2007)}]{AnLittle2007JRSSA}
An, D. and Little, R. J.~A. (2007) Multiple imputation: an alternative to top
  coding for statistical disclosure control.
\newblock \textit{Journal of the Royal Statistical Society, Series A},
  \textbf{170}, 923--940.

\bibitem[{Dimitrakakis et~al.(2017)Dimitrakakis, Nelson, Zhang, Mitrokotsa and
  Rubinstein}]{Dimitrakakis:2017:DPB:3122009.3122020}
Dimitrakakis, C., Nelson, B., Zhang, Z., Mitrokotsa, A. and Rubinstein, B.
  I.~P. (2017) Differential privacy for bayesian inference through posterior
  sampling.
\newblock \textit{J. Mach. Learn. Res.}, \textbf{18}, 343--381.

\bibitem[{Drechsler(2011)}]{Drechsler2011book}
Drechsler, J. (2011) \textit{Synthetic Datasets for Statistical Disclosure
  Control}.
\newblock Springer: New York.

\bibitem[{Drechsler et~al.(2008)Drechsler, Dundler, Bender, Rassler and
  Zwick}]{DrechslerDundlerBenderRasslerZwick2008}
Drechsler, J., Dundler, A., Bender, S., Rassler, S. and Zwick, T. (2008) A new
  approach for disclosure control in the iab establishment panel - multiple
  imputation for a better data access.
\newblock \textit{Advances in Statistical Analysis}, 439--458.

\bibitem[{Drechsler and Hu(2020)}]{DrechslerHu2018}
Drechsler, J. and Hu, J. (2020) Synthesizing geocodes to facilitate access to
  detailed geographical information in large scale administrative data.
\newblock \textit{Journal of Survey Statistics and Methodology}.

\bibitem[{Dwork et~al.(2006)Dwork, McSherry, Nissim and
  Smith}]{Dwork:2006:CNS:2180286.2180305}
Dwork, C., McSherry, F., Nissim, K. and Smith, A. (2006) Calibrating noise to
  sensitivity in private data analysis.
\newblock In \textit{Proceedings of the Third Conference on Theory of
  Cryptography}, TCC'06, 265--284.

\bibitem[{Hu(2019)}]{Hu2019TDP}
Hu, J. (2019) Bayesian estimation of attribute and identification midisclosure
  risks in synthetic data.
\newblock \textit{Transactions on Data Privacy}, \textbf{12}, 61--89.

\bibitem[{Hu et~al.(2018)Hu, Reiter and Wang}]{HuReiterWang2018BA}
Hu, J., Reiter, J.~P. and Wang, Q. (2018) Dirichlet process mixture models for
  modeling and generating synthetic versions of nested categorical data.
\newblock \textit{Bayesian Analysis}, \textbf{13}, 183--200.

\bibitem[{Hundepool et~al.(2012)Hundepool, Domingo-Ferrer, Franconi, Giessing,
  Nordholt, Spicer and de~Wolf}]{SDC2012book}
Hundepool, A., Domingo-Ferrer, J., Franconi, L., Giessing, S., Nordholt, E.~S.,
  Spicer, S. and de~Wolf, P. (2012) \textit{Statistical Disclosure Control}.
\newblock Wiley.

\bibitem[{Karr et~al.(2006)Karr, Kohnen, Oganian, Reiter and
  Sanil}]{KarrKohnenOganianReiterSanil2006}
Karr, A.~F., Kohnen, C.~N., Oganian, A., Reiter, J.~P. and Sanil, A.~P. (2006)
  A framework for evaluating the utility of data altered to protect
  confidentiality.
\newblock \textit{The American Statistician}, \textbf{60}, 224--232.

\bibitem[{Kinney et~al.(2011)Kinney, Reiter, Reznek, Miranda, Jarmin and
  Abowd}]{SynLBD2011}
Kinney, S.~K., Reiter, J.~P., Reznek, A.~P., Miranda, J., Jarmin, R.~S. and
  Abowd, J.~M. (2011) Towards unrestricted public use business microdata: The
  synthetic {L}ongitudinal {B}usiness {D}atabase.
\newblock \textit{International Statistical Review}, \textbf{79}, 363--384.

\bibitem[{Little(1993)}]{Little1993synthetic}
Little, R. J.~A. (1993) Statistical analysis of masked data.
\newblock \textit{Journal of Official Statistics}, \textbf{9}, 407--426.

\bibitem[{Manrique-Vallier and Hu(2018)}]{ManriqueHu2018JRSSA}
Manrique-Vallier, D. and Hu, J. (2018) Bayesian non-parametric generation of
  fully synthetic multivariate categorical data in the presence of structural
  zeros.
\newblock \textit{Journal of the Royal Statistical Society, Series A},
  \textbf{181}, 635--647.

\bibitem[{Quick et~al.(2018)Quick, Holan and Wikle}]{QuickHolanWikle2018JRSSA}
Quick, H., Holan, S.~H. and Wikle, C.~K. (2018) Generating partially synthetic
  geocoded public use data with decreased disclosure risk using differential
  smoothing.
\newblock \textit{Journal of the Royal Statistical Society, Series A},
  \textbf{181}, 649--661.

\bibitem[{Reiter and Mitra(2009)}]{ReiterMitra2009}
Reiter, J.~P. and Mitra, R. (2009) Estimating risks of identification
  disclosure in partially synthetic data.
\newblock \textit{The Journal of Privacy and Confidentiality}, \textbf{1},
  99--110.

\bibitem[{Rubin(1993)}]{Rubin1993synthetic}
Rubin, D.~B. (1993) Discussion statistical disclosure limitation.
\newblock \textit{Journal of Official Statistics}, \textbf{9}, 461--468.

\bibitem[{{Savitsky} and {Toth}(2016)}]{2015arXiv150707050S}
{Savitsky}, T.~D. and {Toth}, D. (2016) {Bayesian estimation under informative
  sampling}.
\newblock \textit{Electronic Journal of Statistics}, \textbf{10}, 1677--1708.

\bibitem[{Snoke et~al.(2018)Snoke, Raab, Nowok, Dibben and
  Slavkovic}]{Snoke2018JRSSA}
Snoke, J., Raab, G.~M., Nowok, B., Dibben, C. and Slavkovic, A. (2018) General
  and specific utility measures for synthetic data.
\newblock \textit{Journal of the Royal Statistical Society, Series A},
  \textbf{181}, 663--688.

\bibitem[{Wei and Reiter(2016)}]{WeiReiter2016}
Wei, L. and Reiter, J.~P. (2016) Releasing synthetic magnitude microdata
  constrained to fixed marginal totals.
\newblock \textit{Statistical Journal of the IAOS}, \textbf{32}, 93--108.

\bibitem[{Williams and Savitsky(2018)}]{WilliamsSavitsky_pairwise2018}
Williams, M.~R. and Savitsky, T.~D. (2018) Bayesian pairwise estimation under
  dependent informative sampling.
\newblock \textit{Electronic Journal of Statistics}, \textbf{12}, 1631--1661.

\bibitem[{Woo et~al.(2009)Woo, Reiter, Oganian and Karr}]{Woo2009JPC}
Woo, M.~J., Reiter, J.~P., Oganian, A. and Karr, A.~F. (2009) Global measures
  of data utility for microdata masked for disclosure limitation.
\newblock \textit{The Journal of Privacy and Confidentiality}, \textbf{1},
  111--124.

\end{thebibliography}


\begin{thebibliography}{4}
\expandafter\ifx\csname natexlab\endcsname\relax\def\natexlab#1{#1}\fi
\expandafter\ifx\csname url\endcsname\relax
  \def\url#1{\texttt{#1}}\fi
\expandafter\ifx\csname urlprefix\endcsname\relax\def\urlprefix{URL: }\fi

\bibitem[{Drechsler(2011)}]{Drechsler2011book}
Drechsler, J. (2011) \textit{Synthetic Datasets for Statistical Disclosure
  Control}.
\newblock Springer: New York.

\bibitem[{Neal(2000)}]{neal:2000}
Neal, R.~M. (2000) Markov chain sampling methods for {D}irichlet process
  mixture models.
\newblock \textit{Journal of Computational and Graphical Statistics},
  \textbf{9}, 249--265.

\bibitem[{Reiter and Raghunathan(2007)}]{ReiterRaghu2007}
Reiter, J.~P. and Raghunathan, T.~E. (2007) The multiple adaptations of
  multiple imputation.
\newblock \textit{Journal of the American Statistical Association},
  \textbf{102}, 1462--1471.

\bibitem[{{Stan Development Team}(2016)}]{Rstan}
{Stan Development Team} (2016) {RStan}: the {R} interface to {Stan}.
\newblock \urlprefix\url{http://mc-stan.org/}.
\newblock R package version 2.14.1.

\end{thebibliography}

\end{document}


\title{\large \bf Supplementary Material for Risk-efficient Bayesian Pseudo Posterior Data Synthesis for Privacy Protection}

\author{Jingchen Hu\footnote{Vassar College, Box 27, 124 Raymond Ave, Poughkeepsie, NY 12604, United States, jihu@vassar.edu.} $\,$ and Terrance D. Savitsky\footnote{U.S. Bureau of Labor Statistics, Office of Survey Methods Research, Suite 5930, 2 Massachusetts Ave NE Washington, DC 20212, United States, Savitsky.Terrance@bls.gov.} and Matthew R. Williams\footnote{National Center for Science and Engineering Statistics, National Science Foundation, 2415 Eisenhower Avenue, Alexandria, VA 22314, United States, mrwillia@nsf.gov}}
\maketitle
\begin{abstract}
This Supplementary Material contains: i) The details of our proposed finite mixutre synthesizer for the CE data; ii) A few toy examples of identifying ``Betty", to illustrate the marginal probability of identification risk in Section 2.1 of the main document; iii) Additional utility results in Section 2.3.1 in the main document; iv) Algorithm for the pairwise risk-weighted pseudo posterior synthesizer in Section 3.1 of the main document; v) Information about the CES survey administrated by the BLS; vi) Pairwise risk-weighted pseudo posterior synthesizer applied to the simulated data, mentioned in Section 3 in the main document; and vii) Two local weight adjustments to our disclosure risk-based weights, offering more flexibility in tuning the utility-risk trade-off.
\end{abstract}

\section{A finite mixture synthesizer for CE data}

Our proposed synthesizer is a flexible, parametric finite mixture synthesizer for a sensitive continuous variable that utilizes available predictors. We describe the synthesizer in the context of the CE data sample to promote ease-of-understanding.  Even though we focus on the CE dataset, this synthesizer is generalizable and widely applicable for synthesizing skewed continuous data.

Let $y_i$ denote the logarithm of the family income for CU $i$, and $\mathbf{X}_i$ be the $R \times 1$ vector including an intercept and the values of predictors of CU $i$. There are $n$ CUs in the sample.
\begin{eqnarray}
	\label{eq:y} y_i \mid \mathbf{X}_i, z_i, \bm{\beta}, \mathbf{\sigma} &\sim& \textrm{Normal}(y_i \mid \mathbf{X}_i^{'}\bm{\beta}^{\ast}_{z_i} , (\sigma^{\ast}_{z_i})^2) \\
	\label{eq:z} z_i \mid \bm{\pi} &\sim& \textrm{Multinomial}(1; \pi_1, \cdots, \pi_K)
\end{eqnarray}

Our finite mixture construction over-specifies the number of mixture components, $K$, to facilitate the flexible clustering of CUs that employ the same generating distribution component for $y$.  Under our modeling setup, we sample locations, $(\bm{\beta}^{\ast}_{k}, (\sigma^{\ast}_{k})^{2})$ and cluster indicators, $z_{i}\in (1,\ldots,K)$, for CU $i$. The cluster indicators, $(z_{i})$ are generated from multinomial draws with cluster probabilities, $\left(\pi_1, \cdots, \pi_K\right)$  in Equation (\ref{eq:z}).

We induce sparsity in the number of clusters selected by the model through our sampling of the $(\pi_{k})$ with,
\begin{align}
\left(\pi_{1},\ldots,\pi_{K}\right) &\sim \mbox{Dirichlet}\left(\frac{\gamma}{K},\ldots,\frac{\gamma}{K}\right), \label{eq:DPprior-pi}\\
\gamma &\sim \textrm{Gamma}(a_{\gamma}, b_{\gamma}).	\label{eq:DPprior-alpha}
\end{align}
Although parametric our model becomes arbitrarily close to a Dirichlet process mixture for an unknown measure, $F$, specified with generating model parameters, $(\bm{\beta}_{k},\sigma_{k}^{2}) \sim F$ in the limit of $K \uparrow \infty$ \citep{neal:2000}. Our parametric formulation denotes a truncated Dirichlet process (TDP).  The $\gamma$ hyperparameter induces sparsity in the number of non-zero cluster probabilities in Equation~(\ref{eq:DPprior-pi}).  Due to its influence on the number of clusters learned by the data, we further place a gamma prior on $\gamma$ in Equation~(\ref{eq:DPprior-alpha}).

We specify an independent and identically-distributed multivariate Gaussian prior distribution for the coefficient locations, $(\bm{\beta}^{\ast}_k)$ in Equation (\ref{eq:prior-beta}). We choose an independent and identically distributed student-t prior for the standard deviation locations, $(\sigma^{\ast}_k)$ in Equation (\ref{eq:prior-sigma}):
\begin{eqnarray}
	\label{eq:prior-beta} \bm{\beta}^{\ast}_k &\iid& \textrm{MVN}_{R}(\mathbf{0}, \mbox{diag}(\bm{\sigma}_{\beta})\times \mathop{\Omega_{\beta}}^{R \times R} \times \mbox{diag}(\bm{\sigma}_{\beta}) ), \\
	\label{eq:prior-sigma} \sigma^{\ast}_{k} &\iid& \textrm{t}(3, 0, 1),
\end{eqnarray}
where the $R \times R$ correlation matrix, $\Omega_{\beta}$, receives a uniform prior over the space of $R \times R$ correlation matrices \citep{Rstan} and each component of $\bm{\sigma}_{\beta}$ receives a student-t prior with $3$ degrees of freedom.

To generate a synthetic family income value for each CU, we first draw sample values of $(\bm{\pi}^{(l)}, \bm{\beta}^{\ast,(l)}, \sigma^{\ast,(l)})$ from the posterior distribution at MCMC iteration $l$. We estimate our TDP mixture model using Stan \citep{Rstan}, after marginalizing out the discrete cluster assignment probabilities, $\mathbf{z}$.  We generate cluster assignments, \emph{a posteriori}, from the full conditional distributions given, $\bm{\pi}^{(l)}$, with,
\begin{equation}
\left(z_{i} \mid (\pi_{k}), (y_{i}), (\beta^{\ast}_{k},\sigma^{\ast}_{k})\right) = \textrm{Multinomial}\left(1; \left[\pi_{1}\phi(y_{i}\mid \mathbf{X}_i^{'}\bm{\beta}^{\ast}_{1},\sigma^{\ast}_{1})\right],\ldots, \left[\pi_{K}\phi(y_{i}\mid \mathbf{X}_i^{'}\bm{\beta}^{\ast}_{K},\sigma^{\ast}_{K})\right]\right),
\end{equation}
where $\phi(\cdot)$ denotes the density function of a normal distribution. We next generate synthetic family income, $\{y_i^{*,(l)}, i = 1, \cdots, n\}$, through a draw from the normally-distributed likelihood given predictor vectors $\{\mathbf{X}_i, i = 1, \cdots, n\}$, and samples of $\mathbf{z}^{(l)}, \bm{\beta}^{\ast,(l)}$ and $\sigma^{\ast,(l)}$, as in Equation (\ref{eq:y}). Let $\mathbf{Z}^{(l)}$ denote a partially synthetic dataset at MCMC
iteration $l$. We repeat the process for $L$ times, creating $L$ independent partially synthetic datasets $\mathbf{Z} = (\mathbf{Z}^{(1)}, \cdots, \mathbf{Z}^{(L)})$.  $L$ is chosen to allow accurate computation of the variance between released datasets for any statistic or parameter of interest estimated by the data analyst from the $L$ synthetic datasets.  The between and within dataset variances are combined using a multiple imputation rule to form a total variance \citep{ReiterRaghu2007, Drechsler2011book}.  

\section{Toy examples for marginal probability of identification risk in Section 2.1}

These toy examples share the goal to identify a person whose information is collected in a survey, named ``Betty", given the known pattern $p$ in which Betty is located, and the true family income, $y_i$.

Recall that the marginal probability of identification risk is defined as,

\begin{equation}
IR^{(l)}_{i}  = \frac{\mathop{\sum}_{h \in M_{p,i}^{(l)}}\mathbb{I}\left(y_{h}^{\ast,(l)}\notin B(y_{i},r)\right)}{\lvert M_{p,i}^{(l)}\rvert} \times T^{(l)}_{i}.
\end{equation}

In the following 4 scenarios, the number of CUs sharing the same pattern as Betty, $p$, is $\lvert M_{p,i}^{(l)}\rvert = 13$. In each of the 4 scenarios, we include the values of $\mathop{\sum}_{h \in M_{p,i}^{(l)}}\mathbb{I}\left(y_{h}^{\ast,(l)}\notin B(y_{i},r)\right)$, $T^{(l)}_{i}$, and $IR^{(l)}_{i}$ in the figure caption. $r$ is the radius of the ball and assumed the same in both scenarios.

\begin{figure}[H]
    \centering
    \subfloat[$IR_i^{(l)} = \frac{10}{13}$]{ \includegraphics[width=0.22\textwidth]{demo1}
    \label{fig:case1}}
    \subfloat[$IR_i^{(l)} = \frac{5}{13}$]{ \includegraphics[width=0.22\textwidth]{demo2}
    \label{fig:case2}}
    \subfloat[$IR_i^{(l)} = 0$]{  \includegraphics[width=0.22\textwidth]{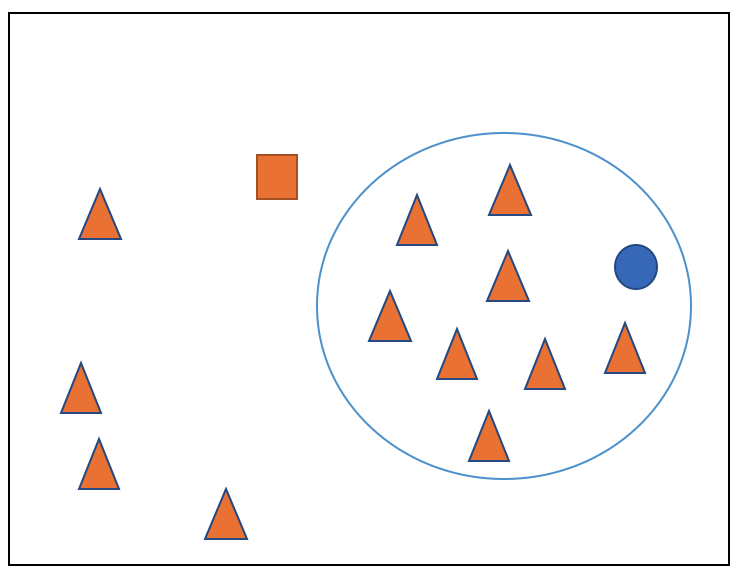}
    \label{fig:case3}}
     \subfloat[$IR_i^{(l)} = 0$]{  \includegraphics[width=0.22\textwidth]{demo4}
    \label{fig:case4}}\\
    \caption{Four cases of Betty's identification disclosure probability, $IR_i^{(l)}$: $|M_{p,i}^{(l)}| = 13$ for all four cases, while $\mathop{\sum}_{h \in M_{p,i}^{(l)}}\mathbb{I}\left(y_{h}^{\ast,(l)}\notin B(y_{i},r)\right)$ is 10, 5, 5, and 10 respectively, and $T^{(l)}_{i} = 1$ is 1, 1, 0, and 0, respectively.}
    \label{fig:cases124}
\end{figure}

Figure \ref{fig:case1} illustrates the case where few records in the pattern are close to Betty's true record value for $y$.  This scenario leads to a relatively higher identification disclosure probability.

Figure \ref{fig:case2} illustrates the case where Betty's true value for $y$ is well covered by many records in the pattern. This scenario leads to a relatively lower probability of identification disclosure.

\begin{figure}[H]
\centering
\includegraphics[width=0.8\textwidth]{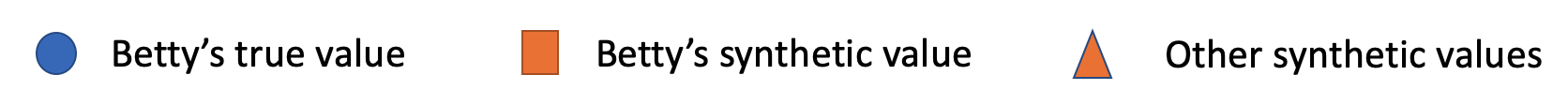}
\end{figure}

Figure \ref{fig:case3} is similar to the case in Figure \ref{fig:case2}, but now Betty's true value for $y$ is not close to her synthesized value, which leads to $T_{i}^{(l)} = 0$ and an identification disclosure probability of 0.

Figure \ref{fig:case4} is similar to the case of Figure \ref{fig:case1}, but now Betty's true value for $y$ is not close to her synthesized value, which leads to $T_{i}^{(l)} = 0$ and an identification disclosure probability of 0. 

\section{Additional utility results in Section 2.3.1}

\begin{table}[H]
\centering
\caption{Table of empirical CDF utility results.}
\begin{tabular}{l | c c}
\hline
 &  $U_m$ (maximum) & $U_a$ (average squared distance)  \\ \hline
 Synthesizer & 0.0151 & 3.6e-05  \\
Marginal &  0.0755 & 0.0018  \\ \hline
\end{tabular}
\label{tab:ecdf_marginal}
\end{table}

\section{Algorithm for pairwise risk-weighted pseudo posterior synthesizer in Section 3.1}
\label{algorithm:pairwise}

We summarize our process in Section 3.1 for generating synthetic data from the pairwise risk-weighted pseudo posterior and evaluating the disclosure probabilities for synthetic data records in Algorithm~\ref{alg:pairwise}. Note that the disclosure risks of synthetic datasets are based on marginal disclosure risk probability in Section 2.1.

\begin{algorithm}[H]
\setstretch{1.1}
\SetAlgoLined
1. Compute $M_{p,(i,j)}$, the set of records in the confidential data sharing the same pattern as records $i$ and $j$ for $i = 1,\ldots,n$ and $j \neq i$ data records.

2. Cast a ball $B(y_i, r)$ with a radius $r$ around the \emph{true} value of record $i$, $y_{i}$, and another $B(y_j, r)$ around $y_{j}$, and count the number of confidential records falling outside the radius $\sum_{h \in M_{p,(i,j)}} \mathbb{I} \left(y_h \notin B(y_j, r) \cap y_h \notin B(y_(j, r) \right)$.

3. Compute the pariwise risk probability, $IR_{i,j}^c \in [0, 1]$ as $IR_{i,j}^c = \sum_{h \in M_{p,(i,j)}} \mathbb{I} \left(y_h \notin B(y_j, r) \cap y_h \notin B(y_j, r) \right) / \lvert M_{p,(i,j)} \rvert$.

4. Formulate by-record weights, $\bm{\alpha}^{pw} = (\alpha_1^{pw}, \cdots, \alpha_n^{pw})$, $\alpha_i^{pw} =  1 - \frac{\sum_{j \neq i, j \in M_{p, i}}IR_{i,j}^c}{|M_{p,i}| - 1}$.

5. Use $\bm{\alpha}^{pw} = \left(\alpha_{1}^{pw},\ldots,\alpha_{n}^{pw}\right)$ to construct the pseudo likelihood from which the pseudo posterior is estimated. 

6. Estimate parameters from the risk-weighted pseudo posterior,
$
p_{\alpha^{pw}}\left(\bm{\theta} \mid \mathbf{y}, \mathbf{X}, \bm{\eta}\right) \propto \left[\mathop{\prod}_{i=1}^{n}p\left(y_{i} \mid \mathbf{X}, \bm{\theta}\right)^{\alpha_{i}^{pw}}\right]p\left(\bm{\theta}\mid \bm{\eta}\right).
$

7. Draw $L$ synthetic datasets, $\bm{Z} = (\bm{Z}^{(1)},\ldots,\bm{Z}^{(L)})$, from the model posterior predictive distribution.

8. Compute the identification disclosure risks on each synthetic dataset $\bm{Z}^{(l)}$, $l \in (1,\ldots,L)$, using $
IR^{(l)}_{i}  = \frac{\mathop{\sum}_{h \in M_{p,i}^{(l)}}\mathbb{I}\left(y_{h}^{\ast,(l)}\notin B(y_{i},r)\right)}{\lvert M_{p,i}^{(l)}\rvert} \times T^{(l)}_{i},
$
where $y_{h}^{\ast,(l)}$ denotes a synthetic data record in synthetic dataset $\bm{Z}^{(l)}$.

9. Take the average of $IR_i^{(l)}$ across $L$ synthetic datasets and use $IR_i = \frac{1}{L}\sum_{l=1}^{L} IR_i^{(l)}$ as the final record-level identification disclosure risk in the synthetic datasets for CU $i$. 
 \caption{Steps to implement the Pairwise risk-weighted synthesizer.}
 \label{alg:pairwise}
\end{algorithm}
\vspace{1mm}

\section{Information about the CES survey in Section 2.4}
\label{sim:data}

The right-skewed data generated under our simulation set-up with a 2-component mixture of negative binomial distribution may represent a total employment variable for business establishments collected from the Current Employment Statistics (CES) survey administered by the BLS. The CES collects total employment and wages from a survey of about $~150,000$ business establishments in the U.S. CES defines a business establishment as ``An establishment is an economic unit, such as a factory, mine, store, or office that produces goods or services. It generally is at a single location and is engaged predominantly in one type of economic activity. Where a single location encompasses two or more distinct activities, these are treated as separate establishments, if separate payroll records are available, and the various activities are classified under different industry codes."\footnote{For more information about the CES: \url{https://www.bls.gov/ces/}.}

The distribution of total employment tends to express a high degree of over-dispersion and skewness due to the large variations in size of employers that mirrors the large variation of the family income variable in the CE data.

\section{Pairwise on the simulated data in Section 3}
\label{pairwise:sim}

We continue our simulation study introduced in Section 2.4. We focus on comparing the risk reduction performance of the Pairwise to the Marginal under the challenging case of $r = 15\%$ where we noted the whack-a-mole phenomenon produced a slight increase in average of by-record risks for the Marginal relative to the unweighted Synthesizer by increasing the risk of moderate risk records in the range of 0.10
- 0.25.

\begin{figure}[H]
  \centering
    \includegraphics[width=0.6\textwidth]{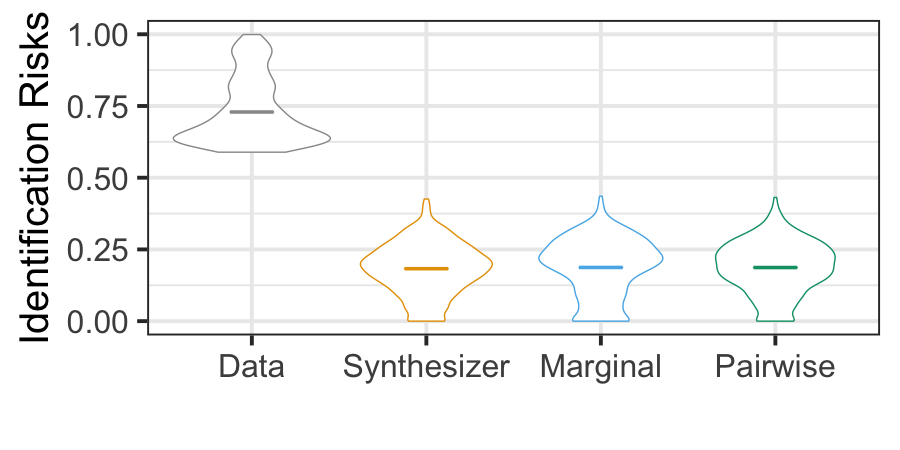}
    \caption{Violin plots of the identification risk probability distributions of the confidential simulated sample, the unweighted Synthesizer, the Marginal synthesizer,  and  the  Pairwise synthesizer. The horizontal bar denotes the mean.}
    \label{fig:sim_risks_marginal_pairwise}
\end{figure}

Figure~\ref{fig:sim_risks_marginal_pairwise} demonstrates that the Pairwise has reduced risks of records in the range of 0.1 - 0.25, rather than increasing them as does the Marginal.   As we saw in the CE application, the Pairwise compresses the by-record risks to achieve risk reduction.

\begin{figure}[H]
    \centering
    \subfloat[Marginal]{   \includegraphics[width=0.2\textwidth]{NBmixture_WAM_marginal_r0p15_diff0p2.png}
      \label{fig:sim_WAM0p2_marginal}}
    \subfloat[Pairwise]{    \includegraphics[width=0.2\textwidth]{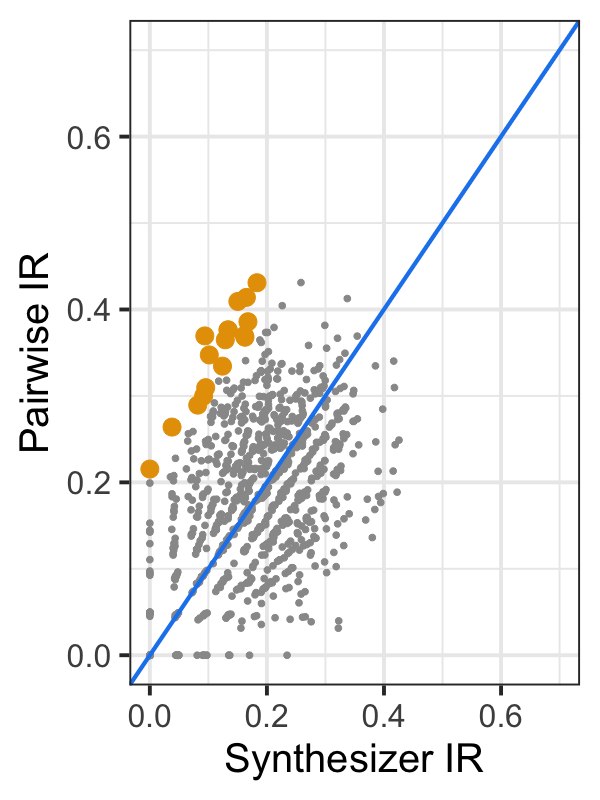}
    \label{fig:sim_WAM0p2_pairwise}}
    \caption{Scatterplot of identification risk of records using the Marginal / Pairwise synthesizer (y-axis) and using the unweighted Synthesizer (x-axis).}
    \label{fig:sim_WAM}
\end{figure}

Figure~\ref{fig:sim_WAM} presents scatterplots of by-record identification risks the Marginal and Pairwise versus the Synthesizer.  This figure confirms the relative improvement in the Pairwise in addressing the whack-a-mole phenomenon since the yellow points representing an increased risk are fewer and the magnitude is reduced.

\begin{table}[H]
\centering
\caption{Table of empirical CDF utility results.}
\begin{tabular}{c | c c}
\hline
 &  $U_m$ (maximum) & $U_a$ (average squared distance)  \\ \hline
 Synthesizer & 0.0451 & 0.0006  \\
Marginal & 0.1088 & 0.0043 \\ 
Pairwise & 0.0378 & 0.0003 \\ \hline
\end{tabular}
\label{tab:ecdfpw}
\end{table}

\begin{figure}[H]
    \centering
    \subfloat[Mean statistic]{   \includegraphics[width=0.45\textwidth]{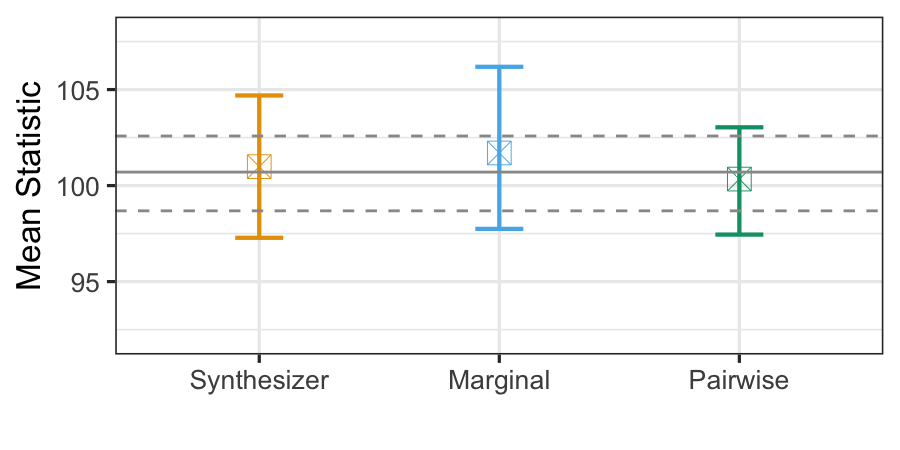}
      \label{fig:sim_utility_mean_pw}}
    \subfloat[Median statistic]{    \includegraphics[width=0.45\textwidth]{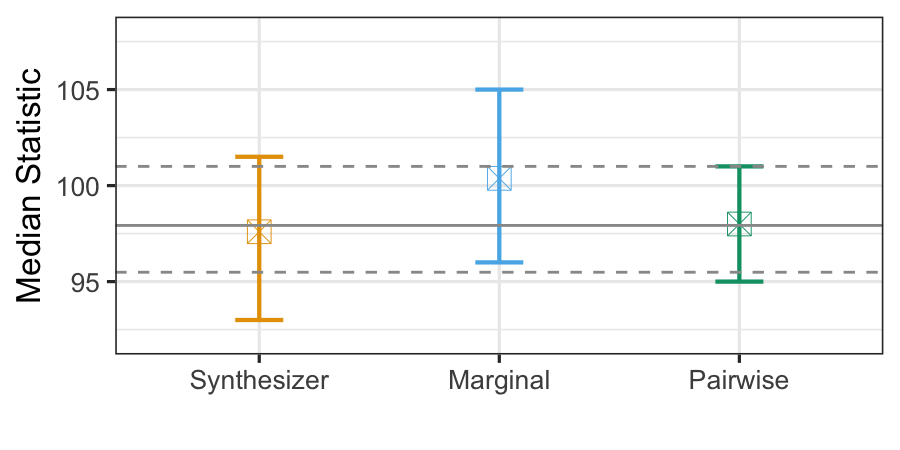}
    \label{fig:sim_utility_median_pw}}
    \caption{Utility plots, comparing point estimates and 95\% confidence interval coverage of Synthesizer, Marginal, and Pairwise (cross in a box represents the mean).  The horizontal solid line and dashed lines represent the point estimate and the 95\% confidence intervals based on the confidential data, respectively.}
    \label{fig:sim_utility_pw}
\end{figure}

Table~\ref{tab:ecdfpw} and Figure~\ref{fig:sim_utility_pw} present analysis-specific and global utility results of the resulting synthetic data from the unweighted Synthesizer, the Marginal and the Pairwise. 
Both utility measures are notably better for the Pairwise than the Marginal due to the mitigation in the whack-a-mole phenomenon.


\section{Two local weight adjustments to tune the utility-risk trade-off}
\label{pairwise:local}
In Section 3.2 of the main document, the two risk-weighted pseudo posterior synthesizers, the Marginal and the Pairwise, have been demonstrated to offer higher privacy protection, compared to the unweighted Synthesizer. The Pairwise gives better control of the overall risk profile and the tail of the record-level identification risk distribution by compression, while maintaining a relatively high level of utility preservation. The Marginal, by contrast, provides better privacy protection than the Synthesizer, but with a bigger compromise on utility.

In this section, we assume that the BLS is satisfied with the privacy protection levels offered by the two risk-weighted pseudo posterior synthesizers, but not yet satisfied with their levels of utility preservation. We propose methods to increase their utility preservation levels, with acceptable loss of the privacy protection. We now proceed to describe the two strategies for achieving a desired utility-risk trade-off balance by the BLS. We focus on the Pairwise due to its superior performance in the trade-off between utility and risk performances, offering notably better risk protection than the Synthesizer for slightly reduced utility. 

\subsection{Methods}
\label{pairwise:local:methods}

The first strategy utilizes a scaling constant $c$ to be applied to the final pairwise probability-based weights in Equation (\ref{eq:pairwiseweights5}). This scaling constant $c$ serves as a tuning parameter for the BLS to control the amount of downweighting of all CUs.  For example, suppose CU $i$ has $\alpha_i^{pw} = 0.5$.  Increasing of $c = 1$ to $c = 1.5$ lifts the final pairwise weight of this CU, $\alpha_i^{pw*}$, from $1 \times 0.5 = 0.5$ to $1.5 \times 0.5 = 0.75$, which translates to a decrease of the amount of downweighting by $0.25$. Increasing the pairwise weight will increase the likelihood contribution of CU $i$ in the corresponding risk-weighted pseudo posterior synthesizer, and is expected to result in higher level of utility preservation.
\begin{equation}
\alpha_i^{pw*} = \textrm{min}(c \times \alpha_i^{pw}, 1)
\label{eq:pairwiseweights5}
\end{equation}

In the limit of increasing $c$ under the setup of Equation (\ref{eq:pairwiseweights5}), each CU receives a weight of 1, which turns the Pairwise risk-weighted pseudo posterior synthesizer to the unweighted synthesizer. We consider the Synthesizer (i.e. weights = 1) as the best scenario of utility preservation. A risk-weighted pseudo posterior synthesizer induces misspecification in the pseudo likelihood. It surgically distorts the portion of the distribution with high identification risks. The induced misspecification translates to less-than-1 weights for some records. Such reduction in weights in turn produces synthetic datasets with lower level of utility preservation.

\begin{figure}
    \centering
    \subfloat[$c = 1, g = 0$]{   \includegraphics[width=0.3\textwidth]{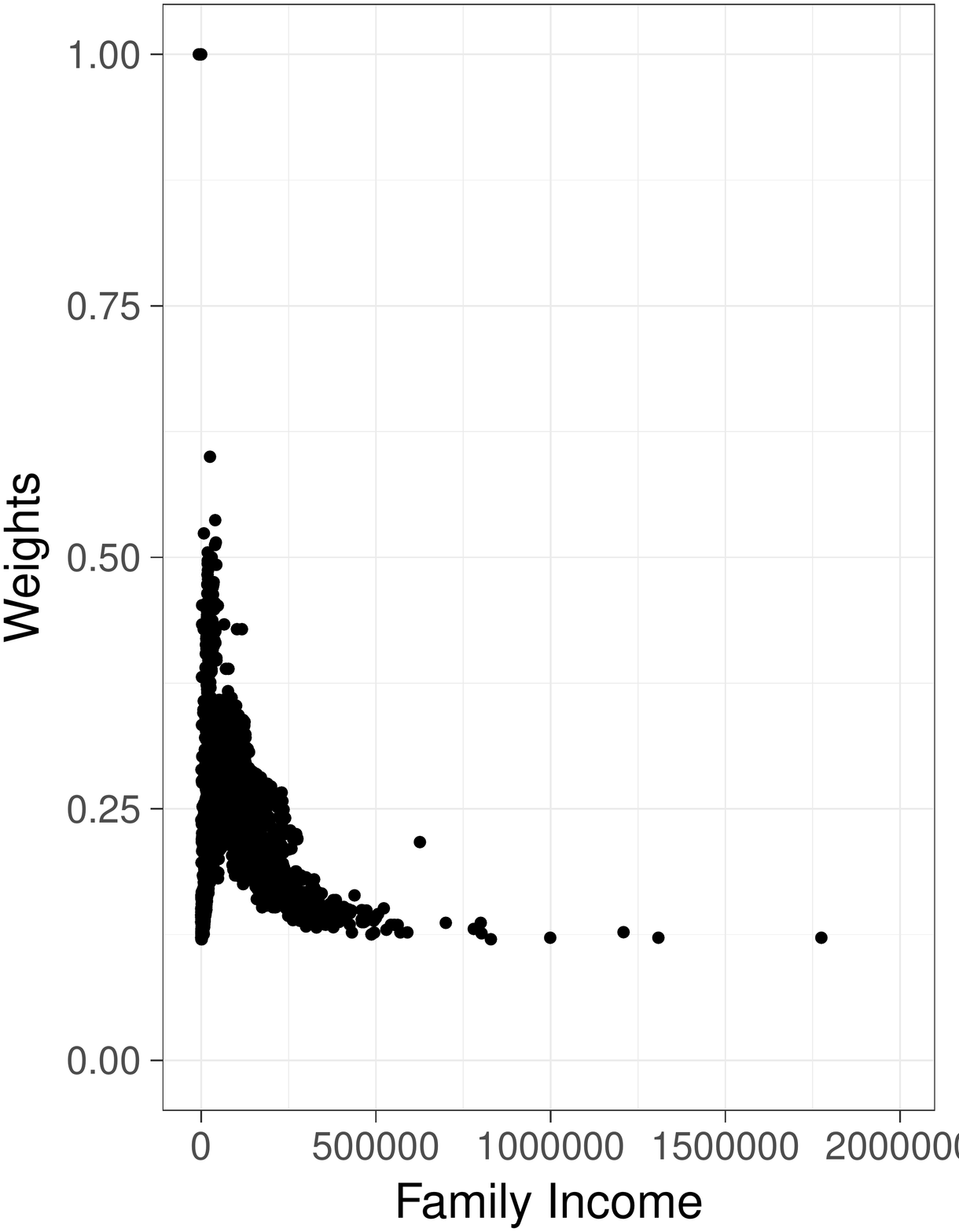}
      \label{fig:pairwiseweightsc1p0}}
    \subfloat[$c = 1.5, g = 0$]{    \includegraphics[width=0.3\textwidth]{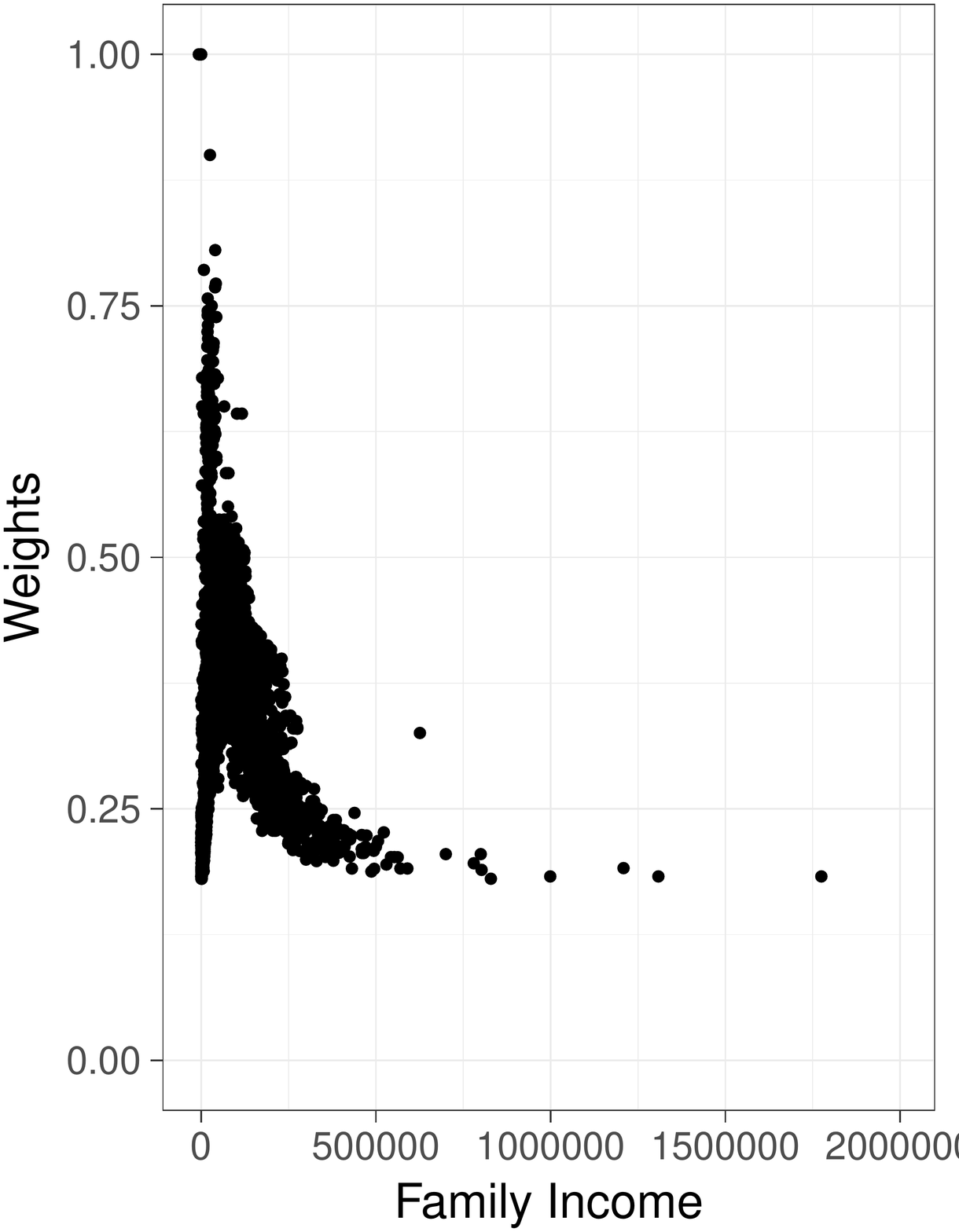}
    \label{fig:pairwiseweightsc1p5}}
    \subfloat[$c = 1, g = 0.1$]{    \includegraphics[width=0.3\textwidth]{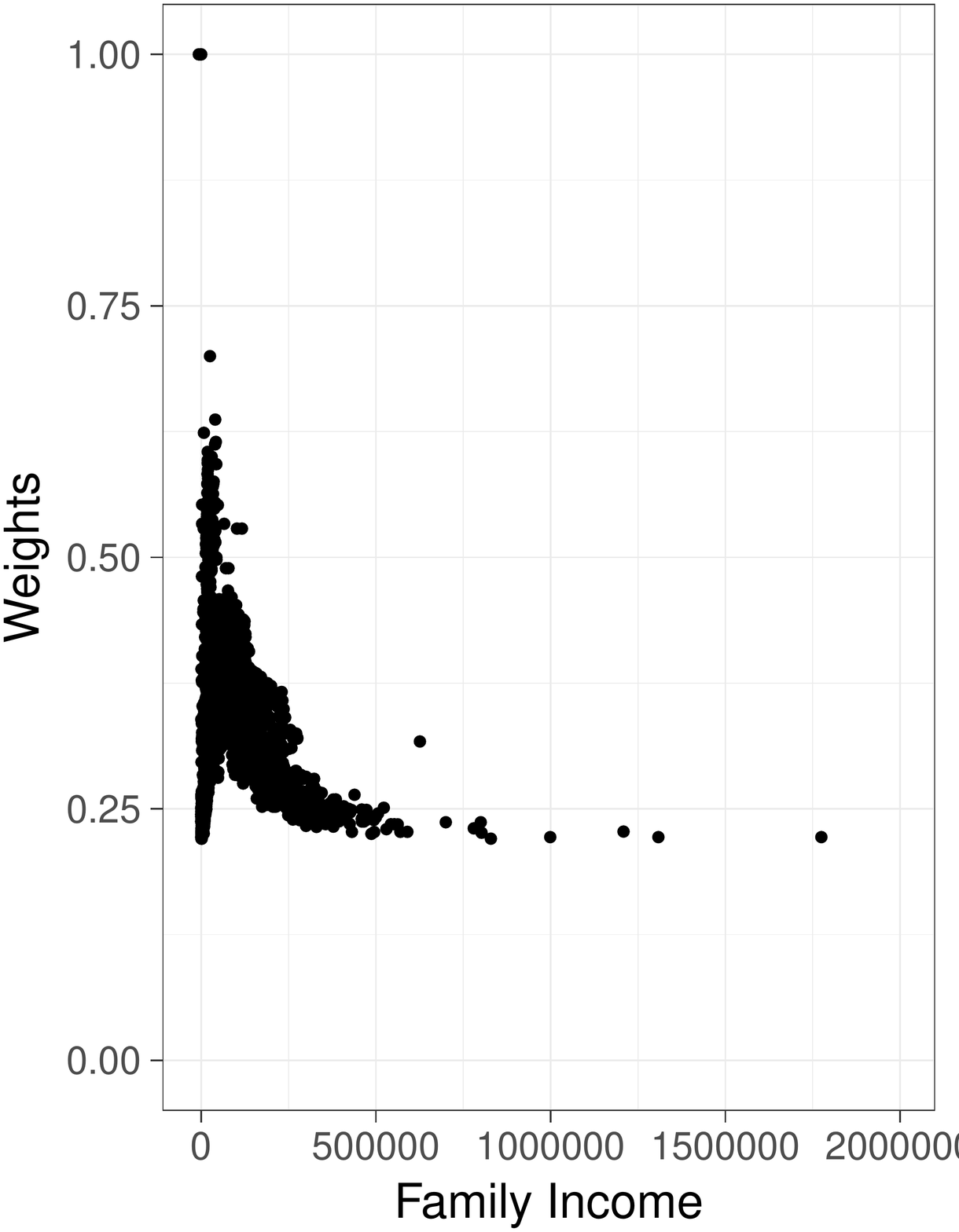}
    \label{fig:pairwiseweightsc1p0g0p1}}
    \caption{Scatter plots of pairwise weights against family income.}
    \label{fig:pairwiseweights_3}
\end{figure}

Yet, it is important to note that the scaling constant $c$ affects the CUs differently. For example, for another CU $j$ with $\alpha_j^{pw} = 0.2$, the increase to $\alpha_j^{pw*}$ is 0.1 when $c$ is increased from $c = 1$ to $c = 1.5$.  By contrast, an increase of $0.25$ occurs for CU $i$ with $\alpha_i^{pw} = 0.5$, producing $\alpha_i^{pw*} = 0.75$.  Tuning $c$ affects all CUs, but to different degrees. Figure \ref{fig:pairwiseweightsc1p5} plots the pairwise weights (y-axis) against family income for all CUs and shows the effects of $c = 1.5$ on the final pairwise weights of all CUs, compared to Figure \ref{fig:pairwiseweightsc1p0} where $c = 1$. A greater-than-1 value of $c$ induces a \emph{stretching} in the final pairwise weights.  A scaling by $c > 1$ affects CUs with higher weights to a greater degree by producing larger magnitude increases of their weights, which is an obvious property of scaling (but, nevertheless, worth noting due to its impact on the identification risk distribution).

Another approach to adjusting the pairwise weights applies a constant with \emph{equal} effect on the final pairwise weights of all CUs. This can be done through adding a constant $g$ in the final pairwise weights construction, as in Equation (\ref{eq:pairwiseweights6}). Figure \ref{fig:pairwiseweightsc1p0g0p1} illustrates the case with $g = 0.1$ while keeping $c = 1$. Compared to Figure \ref{fig:pairwiseweightsc1p0}, the pairwise weight of every CU is increased by 0.1 in Figure \ref{fig:pairwiseweightsc1p0g0p1}, showing that the effect of $g$ is equally applied to the final pairwise weight of each CU, because the additive constant, $g$, \emph{shifts} the distribution of by-record identification risks. A positive $g$ increases the likelihood contribution of all CUs by the same amount, and is expected to result in higher level utility preservation, as every weight is closer to 1 than before.
\begin{equation}
\alpha_i^{pw*} = \textrm{min}(c \times \alpha_i^{pw} + g, 1)
\label{eq:pairwiseweights6}
\end{equation}

It is important to note that when setting $c = 1$ and $g = 0$ in Equation (\ref{eq:pairwiseweights6}), we obtain the pairwise risk-weighted pseudo posterior synthesizer in the application in Section 3.2 of the main document. Figure \ref{fig:pairwiseweightsc1p0} illustrates this basic setup, and we observe how utilizing the pairwise identification risk probabilities to construct weights produces a selective downweighting of CUs as compared to the unweighted Synthesizer where every CU receives a weight of 1. The majority of the CUs under the Pairwise shown in Figure \ref{fig:pairwiseweightsc1p0} receive weights around 0.25, with just a few CUs having weights of 1 and many CUs with large or extremely large family income values having weights as low as 0.1. We utilize the pairwise weights in order to achieve a relatively large or \emph{global} effect on the utility-risk trade-off.   The further tuning of the pairwise weights using $c$ and $g$ are designed to induce a relatively small or \emph{local} effect on the resulting by-record weights to allow a more precise setting of the utility-risk trade-off balance sought by the statistical agency.  A greater-than-1 value of $c$ has a stretching effect on the distribution of the pairwise weights, while a positive value of $g$ induces an upward shift in the weight distribution. It is possible to tune $c$ and $g$ at the same time. For simplicity and illustration purpose, we evaluate tuning only one of these two parameters. 

It bears mention that one may set $c < 1$ and $g < 0$ in the case the statistical agency desires to locally adjust the risks further downward.  We decide to focus on the $c > 1$ and $g > 0$ cases of allowing a bit more risk to improve utility because it is the situation faced by the BLS on the CE sample data as they were willing to accept more disclosure risk for better utility. We now turn to the utility and identification risk profiles of the Pairwise risk-weighted pseudo posterior synthesizers with these local weights adjustments.

\subsection{Results}
\label{pairwise:local:results}

For the Pairwise risk-weighted pseudo posterior synthesizers, we consider three variations of the final pairwise weights in Equation (\ref{eq:pairwiseweights6}): i) $c = 1, g = 0$; ii) $c = 1.5, g = 0$; and iii) $c = 1, g = 0.1$. The choice of the radius $r$, the assumption of intruder's knowledge, and the configurations of the synthesis, stay the same as in Section 3.2 of the main document. As before, we utilize the results of the confidential CE sample and the unweighted Synthesizer for context and comparison.

\begin{figure}[H]
    \centering
    \subfloat[Mean statistic]{ \includegraphics[width=0.45\textwidth]{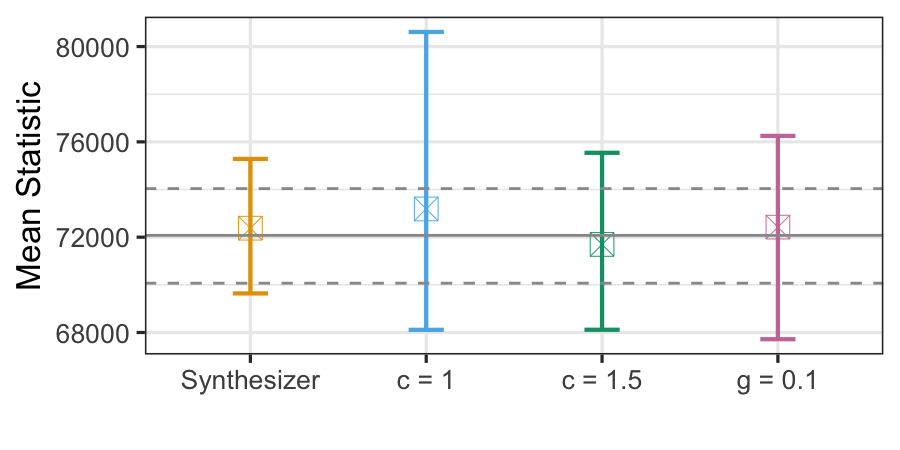}
    \label{fig:data_utility_mean_s}}
    \subfloat[Median statistic]{ \includegraphics[width=0.45\textwidth]{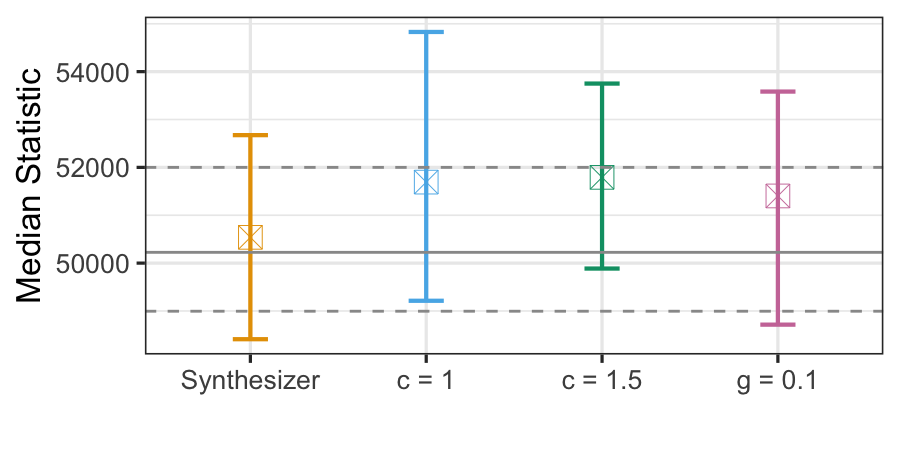}
    \label{fig:data_utility_median_s}}\\
     \subfloat[90\% quantile statistic]{  \includegraphics[width=0.45\textwidth]{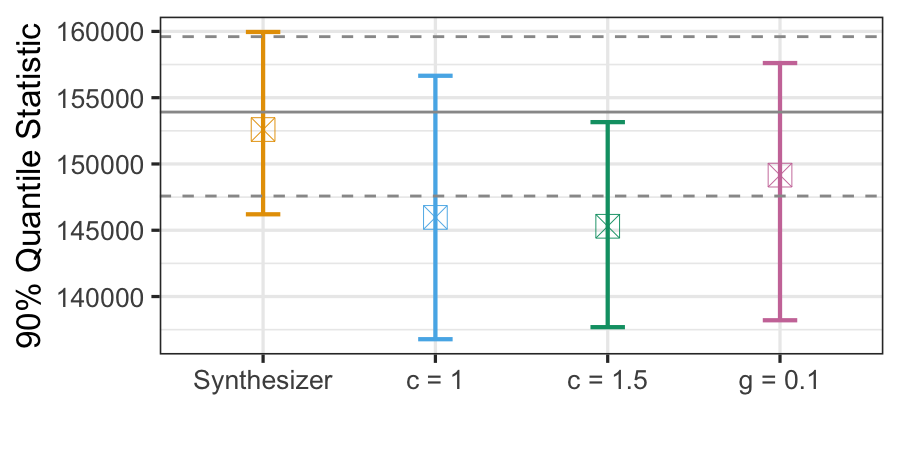}
    \label{fig:data_utility_q90_s}}
     \subfloat[Earner 2 regression coefficient]{  \includegraphics[width=0.45\textwidth]{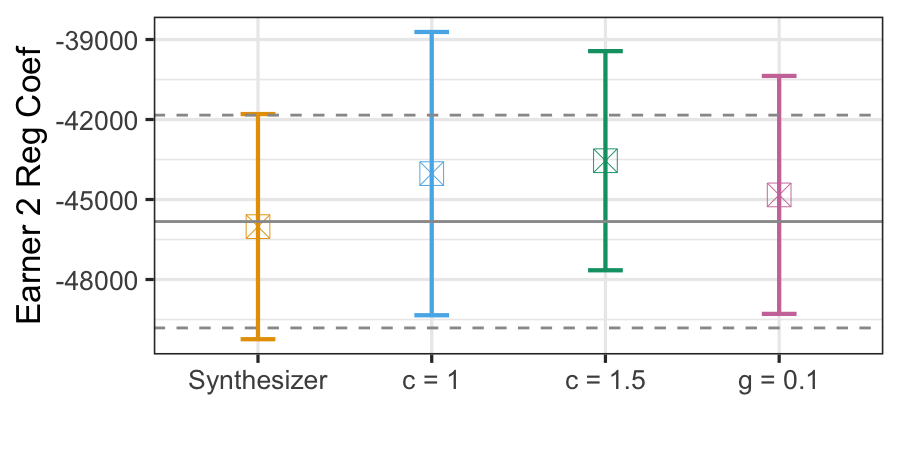}
    \label{fig:data_utility_earner2_s}}\\
    \caption{Utility plots, comparing point estimates and 95\% confidence interval coverage of Synthesizer and different Pairwise risk-weighted pseudo posterior synthesizers  (cross in a box represents the mean). The horizontal solid line represents the point estimate, and the horizontal dashed lines represent the 95\% confidence intervals based on the original data.}
    \label{fig:data_utility_s}
\end{figure}

With increased weights through greater-than-1 values of $c$ or positive values of $g$, we expect to see increased utility preservation by the Pairwise risk-weighted pseudo posterior synthesizers with these local weights adjustments. As before, we perform the same analysis-specific utility evaluations and present the results in Figure \ref{fig:data_utility_s}. These utility results suggest that setting $g = 0.1$ offers improvement in all utility measures, compared to setting $c = 1$. Setting $c = 1.5$, on the other hand, offers smaller utility improvement.


The comparisons of differences between the empirical distributions of the synthesizer and the Pairwise under scaling and shifting also show smaller maximum ($U_{m}$) and average ($U_{a}$) pairwise distances for shifting by $g = 0.1$ than scaling by $c = 1.5$.

\begin{table}[H]
\centering
\begin{tabular}{c | c c}
\hline
 &  $U_m$ (maximum) & $U_a$ (average squared distance) \\ \hline
Synthesizer & 0.0151 & 3.6e-05  \\
Pairwise $c = 1, g = 0 $ & 0.0356 & 0.0004 \\ 
Pairwise $c = 1.5, g = 0 $ & 0.0382 & 0.0004 \\ 
Pairwise $c = 1, g = 0.1 $ & 0.0286 & 0.0002 \\ \hline
\end{tabular}
\caption{Table of empirical CDF utility results.}
\label{tab:cgecdf}
\end{table}

\begin{figure}
\centering
\includegraphics[width=0.5\textwidth]{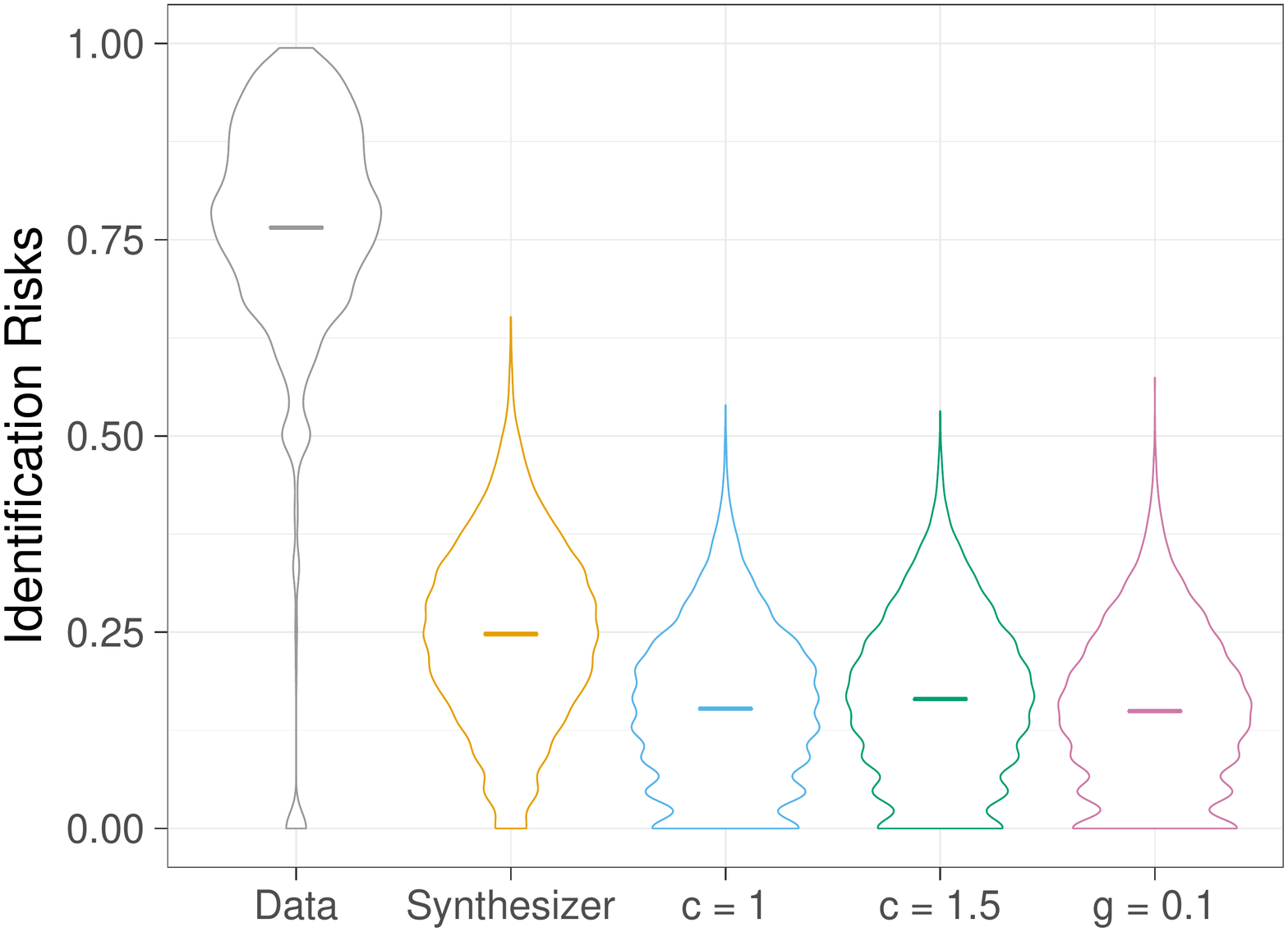}
\caption{Violin plots of the identification risk profiles of the confidential CE sample, the unweighted Synthesizer, and the three pairwise risk-weighted pseudo posterior synthesizers.}
\label{fig:IRpairwise}
\end{figure}

To find out whether such utility improvement comes at a price of reduced privacy protection, we create violin plots to show the identification risk probability distributions in Figure \ref{fig:IRpairwise}. The violin plots show different impacts on identification risks when setting $c = 1.5$ or $g = 0.1$, compared to when setting $c = 1$. Increasing $c$ slightly increases the average identification risks (the horizontal bar), while producing a slightly shorter tail, indicating a better control of the maximum identification risks. Increasing $g$, on the other hand, keeps a similar average of  identification risks, while producing a longer tail, indicating a worse control of the maximum identification risks. Both $c > 1$ and $g > 1$ provide higher privacy protection compared to the unweighted Synthesizer, in terms of the average and the tail.

In summary, setting $c = 1.5$ or $g = 0.1$ for the Pairwise risk-weighted pseudo posterior synthesizers offers improvement in utility. Tuning a positive $g$ offers higher utility improvement, with a higher price of privacy protection reduction. Tuning a greater-than-1 $c$ offers reasonable utility improvement, producing slightly higher average identification risks and slightly smaller maximum identification risks. Depending on the microdata release policy set by the BLS, the proposed local weights adjustments could help the BLS to achieve their desired utility-risk trade-off balance when disseminating synthetic datasets through the pairwise risk-weighted pseudo posterior synthesizers.

\bibliography{CEbib}